\documentclass[journal]{IEEEtran}

\usepackage{amsfonts}
\usepackage{amsmath}
\usepackage{amssymb}
\usepackage{graphicx}
\usepackage{mathtools}
\usepackage{adjustbox}
\usepackage{hyperref}
\hypersetup{colorlinks,allcolors=black}
\usepackage{comment}
\usepackage{cite} 

\usepackage[caption=false, font=footnotesize]{subfig} 

\usepackage{amsthm} 

\usepackage{soul}  

\usepackage[normalem]{ulem}
\newtheorem{theorem}{Theorem}

\newtheorem{proposition}{Proposition}

\usepackage{hyperref}

\usepackage[makeroom]{cancel} 

\usepackage{xcolor} 

\usepackage{algorithm}
\usepackage{algorithmicx}
\usepackage{algpseudocode}

\usepackage{tikz}
\usetikzlibrary{shapes.multipart}

\usetikzlibrary{circuits.logic.US, positioning}
\usetikzlibrary{shapes.geometric, arrows}

\begin{document}
\title{A New Metric Function for SC-based Polar Decoders: Polarization, Pruning, and Fast Decoders}
\author{Mohsen~Moradi\textsuperscript{\href{https://orcid.org/0000-0001-7026-0682}{\includegraphics[scale=0.06]{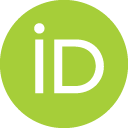}}},
Hessam~Mahdavifar\textsuperscript{\href{https://orcid.org/0000-0001-9021-1992}{ \includegraphics[scale=0.06]{figs/ORCID}}}
\thanks{An earlier version of this paper was presented in part at the 2025 IEEE International Symposium on Information Theory.}
\thanks{The authors are with the Department of Electrical \& Computer Engineering, Northeastern University, Boston MA-02115, USA (e-mail: m.moradi@northeastern.edu, h.mahdavifar@northeastern.edu).}
\thanks{This work was supported by NSF under Grant CCF-2415440 and the Center for Ubiquitous Connectivity (CUbiC) under the JUMP 2.0 program.}%
}

\maketitle
\begin{abstract} 
In this paper, we propose a method to obtain the optimal metric function at each depth of the polarization tree through a process we call \textit{polarization} of the metric function.
This polarization process generates an optimal metric at intermediate levels of the polarization tree, which can be applied in \textit{fast} successive-cancellation-based (FSC) and SC list-based (FSCL) decoders—decoders that partially explore the binary tree representation. 
\textbf{We prove that at each step of the polarization tree, the expected value of the metric function random variable is the mutual information of the corresponding channel, while its variance equals the varentropy of the channel—two parameters that are particularly relevant in finite block-length regimes. 
Additionally, we show that after polarization, the variances of the bit metrics approach zero for binary-input discrete memoryless channels (BI-DMCs). 
Moreover, we provide an estimate for calculating the variance of the binary-input additive white Gaussian noise (BI-AWGN) channel.}
We introduce a list-pruning strategy for FSCL decoding that retains only the paths whose metric values are close to the average. 
As a result, our method significantly reduces the number of required sorting operations in FSCL-based decoding algorithms.
We also derive an upper bound, as a function of the polarized channel varentropy, on the probability that the distance between a bit-metric random variable and the bit-channel mutual information exceeds a given threshold. 
Leveraging this result, we further propose a varentropy-based list-pruning strategy for the SCL (VPSCL) decoding algorithm that adapts to the varentropy of the corresponding bit-channel.
Our proposed pruning strategy also benefits stack decoding (VPStack) by discarding partial paths and avoiding unnecessary extensions.

\end{abstract}


\section{Introduction}
\IEEEPARstart{P}{olar} codes, introduced by Ar{\i}kan \cite{arikan2009channel}, represent the first class of codes proven to achieve the capacity of memoryless symmetric channels with explicit constructions and low-complexity encoding and decoding. As an improvement to polar codes, polarization-adjusted convolutional (PAC) codes \cite{arikan2019sequential} are a new class of linear codes that demonstrate error-correction performance close to theoretical bounds in the short block-length regime \cite{arikan2019sequential, moradi2020performance}. The main new ingredient of PAC codes is the use of convolutional codes as a pre-transformation that improves upon the weight spectrum of polar codes \cite{li2019pre, moradi2024polarization}. PAC codes are also equivalent to post-transforming polar codes with certain cyclic codes \cite{moradi2024polarization}.

PAC codes can be decoded using the Fano algorithm \cite{fano1963heuristic}.
The Fano algorithm is a sequential decoding method that follows a single path on the decoding tree while accumulating a certain metric function.
Fano-type decoders typically have a varying complexity, which can also become exponentially expensive, which is not appealing from a practical perspective \cite{moradi2023application}. The successive cancellation list (SCL) decoding of polar codes \cite{tal2015list} is also adapted for decoding PAC codes \cite{yao2021list, rowshan2021polarization}. 
This approach requires an increase in the memory size by a factor of \(L\), where \(L\) is the maximum list size, compared to successive cancellation (SC) or Fano decoding. Further, to achieve a performance similar to that of the Fano algorithm, a relatively large list size is required. 
A fundamental parameter in all SC-type decoders for polar codes and PAC codes is the decoding path metric that serves as a measure for the reliability of the corresponding decoding path. The path metric is incrementally updated as the SC-type decoder progresses, where the updates often occur only at the last level of the polarization binary tree, where information bits are decoded.

In this paper, we introduce a method to derive the optimal metric function for each level of the polarization tree, a process that we refer to as the polarization of the metric function. 
We demonstrate that, on average, the values of the metric functions at each level of the polarization tree correspond to the polarized mutual information for the correct branch and a negative value for the incorrect branch. Note that, in the finite blocklength regime, the maximum coding rate for a target probability of error is characterized by the channel capacity as the first-order term, a second-order term that depends on the channel dispersion, together with higher-order terms that vary depending on the channel model \cite{polyanskiy2010channel}. In this work, we establish a connection between the polarization of the optimal metric function and these fundamental channel parameters. In particular, we use these connections to show that for a BI-DMC the variance (varentropy) of the proposed metric function approaches zero in the channel polarization process. We further expand the theoretical analysis of the new metric function, and establish an upper bound on the probability that the difference between the random variable representing the metric function and the polarized mutual information exceeds a given threshold value. Then we conclude that the bit-metric values converge to the polarized mutual information in probability.


An immediate application of the proposed metric function is that it can be utilized in deployment of \textit{fast} SC (FSC) and SCL (FSCL) decoders, i.e., decoders that opt to \textit{prune} special nodes, such as rate-1 and rate-0 nodes, in the binary tree representation for more efficient implementation \cite{alamdar2011simplified, el2017relaxed}. 
Our new method addresses this by providing an optimal metric function that operates at the intermediate levels of the polarization tree, which is essential to tree-pruning techniques. 
Our method provides a framework to apply any tree-pruning technique developed for PAC codes. This includes various special nodes considered including repetition (REP) nodes and single parity-check (SPC) nodes \cite{sarkis2014fast}, as well as the so-called Type-I nodes \cite{hanif2017fast}.  

Furthermore, leveraging our theoretical upper bounds, we propose a list-pruning strategy by specifying a threshold on the decoding metric to eliminate candidates from the list. The proposed threshold is a function of the polarized varentropy. A well-polarized channel has a smaller variance, thus, the absolute value of the list-pruning threshold can be smaller, enabling the SCL decoding algorithm to distinguish between correct and incorrect paths often without the need for sorting operations. This addresses another major bottleneck in the implementation of FSCL decoding, that is, the need to sort at each level of the polarization tree when the number of bifurcated paths exceeds \(L\), in order to identify the best \(L\) paths. 
For instance, for a high-rate PAC\((128,99)\) code constructed using the Reed-Muller rate profile, SCL decoding with a list of size \(32\) delivers error correction performance on par with that of the Fano algorithm. 
In comparison, FSCL decoding requires visiting only \(28\) nodes of the polarization tree—each corresponding to a time step—which represents a substantial reduction from the \(254\) time steps required by conventional SCL decoding.
Furthermore, our approach cuts the number of sorting operations of FSCL algorithm to about \(33\%\) of the previous requirement, which further enhances latency performance at a high $E_b/N_0$ regime.
For a PAC\((1024,512)\) code constructed using the polar rate profile, the reduction in sorting operations for the FSCL algorithm is even more substantial, decreasing to about \(5\%\) of the previous requirement.

Stack decoding~\cite{zigangirov1966some, jelinek1969fast} is a sequential decoding algorithm that explores each path on the code tree at most once and requires sorting the paths in the stack at each decoding step.  
We have extended our pruning strategy to stack decoding (referred to as VPStack) to enable early discarding of unlikely paths during decoding.  
Numerical results for PAC$(128,64)$ demonstrate that VPStack decoder achieves error-correction performance comparable to that of the conventional stack decoder.
Moreover, at \(E_b/N_0 = 3.5\) dB, VPStack reduces the number of paths stored in the stack to approximately 12\% of those in conventional stack decoding.

As another application, the new decoding metric can be used to bridge the theoretical analyses of convolutional codes (CCs) to polar and PAC codes, providing a new perspective for studying these codes. In convolutional codes (CCs), the decoder employs a tree search algorithm and utilizes an optimal metric function to determine the path that is closest to the transmitted codeword. Many theoretical analyses of CCs rely on the assumption that the bit metric function variables are independent and identically distributed (i.i.d.) for each channel use \cite[Ch. 6]{gallager1968information}, \cite[Ch. 6]{jacobs}. In a recent work, inspired by the theoretical analyses of CC codes, the authors have leveraged the properties of the new decoding metric to upper bound the computation distribution of polar-like codes \cite{moradi2024pac}. 

The main contributions of this work are as follows.
\begin{itemize}
    \item We introduced the polarization of the metric function and showed that the resulting metric random variables become independent and identically distributed (i.i.d.) across the polarized channels.
    \item We established a novel connection between the decoding algorithm’s metric function and the channel’s varentropy.
    \item We proposed an efficient approximation method for computing the varentropy of the binary-input AWGN (BI-AWGN) channel.
    \item We developed a varentropy-based list pruning technique for both SCL (VPSCL) and stack (VPStack) decoding algorithms, which effectively identifies and discards incorrect paths to reduce computational complexity.
\end{itemize}

The remainder of this paper is organized as follows. 
Sections \ref{sec: Polar_BG}, and \ref{sec: PAC_BG} provide brief overviews of polar codes, RM codes, and PAC codes. 
In Section \ref{sec: MetricPolarization}, we introduce the concept of metric polarization to derive an optimal metric function for each level of the polarization tree. 
Section \ref{sec: MetricEstimation} presents the proposed average metric function for FSCL decoding and compares it with sample average metric function values across different code rates and lengths. 
Section \ref{sec: NumericalResults} offers simulation results. 
Finally, Section \ref{sec: Conclusion} concludes the paper.

\section{Preliminaries} 

\subsection{Notation Convention}

Vectors and matrices are represented in this paper by boldface letters. 
All operations are performed in the binary field $\mathbb{F}_2$. 
For a vector $\mathbf{u} = (u_1, u_2,\ldots,u_N)$,
subvectors $(u_1, \ldots, u_i)$ and $(u_i, \ldots, u_j)$ are denoted by $\mathbf{u}^i$ and $\mathbf{u}_i^j$, respectively.
For a given set $\mathcal{A}$, the subvector $\mathbf{u}_{\mathcal{A}}$ includes all elements $u_i$ where $i$ belongs to the set $\mathcal{A}.$

\subsection{Polar and Reed-Muller Codes}
\label{sec: Polar_BG}

A binary input discrete memoryless channel (BI-DMC) with an arbitrary output alphabet $\mathcal{Y}$ is denoted by $W: \mathcal{X} \longrightarrow \mathcal{Y}$. 
$W(y|x)$ represents the probability of a channel transition, where $x \in \mathcal{X} = \{0,1\}$ and $y \in \mathcal{Y}$. 
By selecting a certain subset of the rows of $\mathbf{F}_N \triangleq \mathbf{F}^{\otimes n}$, which is the $n$-th Kronecker power of $\mathbf{F} = \begin{bsmallmatrix} 1 & 0 \\ 1 & 1 \end{bsmallmatrix}$ with $n = \log_2 N$, the generator matrix of polar codes is obtained. Such a process for selection of a submatrix of $\mathbf{F}_N$, referred to as the rate profiling problem of polar codes, involves identifying a set $\mathcal{A}$ to embed the information (data)
bits in the data carrier vector.

The information vector $\mathbf{d}^K$ of length $K$ can be inserted into the vector $\mathbf{u}^N$ as $\mathbf{u}_{\mathcal{A}} = \mathbf{d}^K$ and $\mathbf{u}_{\mathcal{A}^{c}} = \mathbf{0}$ for an $(N, K, \mathcal{A})$ polar code with $N = 2^n$, where $\mathcal{A} \subseteq \{1,2,\dots,N\}$ corresponds to the rate profile of the code. 
For symmetric channels, the frozen bits $\mathbf{u}_{\mathcal{A}^c}$ are assigned to all zeros, where the complementary set $\mathcal{A}^c$ denotes the frozen bit set. 
Then, the encoding is performed as $\mathbf{x}^N = \mathbf{u}^N \mathbf{F}_N$. 

As a well-studied family of linear block codes, Reed-Muller (RM) codes benefit from their specific and universal construction (i.e., not channel dependent), favorable weight spectrum, and appealing structural properties \cite{muller1954application, reed1953class}. RM codes can be also viewed as codes in the general family of polar codes. More specifically,
an RM$(r,m)$ code, with length $N = 2^m$ and dimension $K = \binom{m}{0} + \binom{m}{1} + \cdots + \binom{m}{r}$, where $\binom{m}{t}$ is a binomial coefficient, can be constructed by selecting all $K$ rows of the matrix $\mathbf{F}_N$ that have Hamming weights greater than or equal to $2^{m-r}$. 
The selection of row indices is a key distinction between RM codes and that of polar codes driven by their successive cancellation (SC)-type decoders. 
Additionally, the code dimension $K$ in an RM code can take on $m+1$ distinct values, whereas in polar codes $K$ can be arbitrarily set as long as $1 \leq K \leq N$. 
In this paper, we use the notation RM$(N, K)$ instead of the more conventional RM$(r,m)$ notation.


\tikzstyle{block} = [rectangle, minimum width=1.5cm, minimum height=1cm, text centered, draw=black, align=center]
\tikzstyle{arrow} = [thick, ->, >=stealth]

\begin{figure}[htbp]
\centering
\resizebox{\columnwidth}{!}{
\begin{tikzpicture}[node distance=1.5cm, every node/.append style={font=\footnotesize}]
\node (Insertion) [block] {Data \\ Insertion};
\node (conv_enc) [block, right of=Insertion, xshift=1cm] {Convolutional \\ Encoder};
\node (p_enc) [block, right of=conv_enc, xshift=1cm] {Polar \\ Mapper};
\node (ch) [block, right of=p_enc, xshift=.25cm, yshift=-1.25cm] {Channel};
\node (p_dec) [block, below of=p_enc, yshift=-1cm] {Polar \\ Demapper};
\node (conv_dec) [block, below of=conv_enc, yshift=-1cm] {Tree \\ Search};
\node (Extraction) [block, below of=Insertion, yshift=-1cm] {Data \\ Extraction};

\draw [arrow] ([xshift=-1cm]Insertion.west) -- (Insertion.west) node[midway, above] {$\mathbf{d}$}; 
\draw [arrow] (Insertion) -- (conv_enc) node[midway, above] {$\mathbf{v}$};
\draw [arrow] (conv_enc) -- (p_enc) node[midway, above] {$\mathbf{u}$};
\draw [arrow] (p_enc.east) -| (ch.north) node[midway, above] {$\mathbf{x}$};
\draw [arrow] (ch) |- (p_dec) node[midway, below] {$\mathbf{y}$};
\draw [arrow] (p_dec) -- (conv_dec) node[midway, above] {metric};
\draw [arrow] (conv_dec) -- (Extraction) node[midway, above] {$\hat{\mathbf{v}}$};
\draw [arrow] (conv_dec.south) -- ++(0, -.5cm)  -| (p_dec.south)  node[pos=0.75, above, yshift=-0.3cm, xshift=-0.3cm] {$\hat{\mathbf{u}}^{i-1}$}; 
\draw [arrow] (Extraction.west) -- ++(-1cm,0) node[midway, above] {$\hat{\mathbf{d}}$}; 

\end{tikzpicture}
}
\caption{Flowchart of PAC coding scheme.}
\label{fig: flowchart}
\end{figure}

\subsection{PAC Coding Scheme} 
\label{sec: PAC_BG}
For an $(N, K, \mathcal{A}, \mathbf{T})$ PAC code, Fig. \ref{fig: flowchart} illustrates a block diagram of the PAC coding scheme. Here, $N$ and $K$ represent the code length and data length, respectively. The set $\mathcal{A}$ defines the code rate profile, and $\mathbf{T}$ is a rate-one convolutional code generator matrix. More specifically, it is an upper-triangular Toeplitz matrix obtained from a connection polynomial $\mathbf{p}(x) = p_m x^{m} + \cdots + p_1 x + p_0$, where $p_0 = p_m = 1$, and is given by

\begin{equation*}
\mathbf{T} = 
\begin{bmatrix}
 p_0    & p_1    &  p_2   & \cdots & p_m    & 0      & \cdots & 0      \\
 0      & p_0    & p_1    & p_2    & \cdots & p_m    &        & \vdots \\
 0      & 0      & p_0    & p_1    & \ddots & \cdots & p_m    & \vdots \\
 \vdots & 0      & \ddots & \ddots & \ddots & \ddots &        & \vdots \\
 \vdots &        & \ddots & \ddots & \ddots & \ddots & \ddots & \vdots \\
 \vdots &        &        & \ddots & 0      & p_0    & p_1    & p_2    \\
 \vdots &        &        &        & 0      & 0      & p_0    & p_1    \\
 \vdots & \cdots & \cdots & \cdots & \cdots & 0      & 0      & p_0    
\end{bmatrix}.
\end{equation*}
In this paper, we use the connection polynomial $\mathbf{p}(x) = x^{10} + x^9 + x^7 + x^3 + 1$ for our simulations \cite{moradi2020performance}.

Same as in polar codes, the source vector $\mathbf{d}^K$ is assumed to be randomly generated from a uniform distribution over $\{0,1\}^K$. The data insertion block, based on the rate profile $\mathcal{A}$, converts these $K$ bits into a data carrier vector $\mathbf{v}^N$, establishing a code rate $R = K / N$ ($\mathbf{v}_{\mathcal{A}} = \mathbf{d}^K$ and $\mathbf{v}_{\mathcal{A}^c} = \mathbf{0}$). 
Subsequently, the convolutional encoder from $\mathbf{v}^N$ generates $\mathbf{u}^N = \mathbf{v}^N \mathbf{T}$. 
This step imposes a constraint on each element $u_j$ in $\mathbf{u}^N$, dependent on at most $m$ prior bits. 
The polar mapper then encodes $\mathbf{u}^N$ into $\mathbf{x}^N = \mathbf{u}^N \mathbf{F}_N$. 
During decoding, the SCL decoder estimates $\hat{v}_i$. 
Finally, the decoded data $\hat{\mathbf{v}}^N$ allows extraction of the $K$-bit information based on $\mathcal{A}$. 
Instead of an SCL decoder, other tree search algorithms can also be used.

\subsection{FSC, FSCL and Metric Functions}

Fano, successive cancellation (SC) and successive cancellation list (SCL) decoding of polar and PAC codes involve traversing a polarization tree, corresponding to the channel polarization process, which is a complete tree of depth \(n = \log_2(N)\) for a polar or PAC code of length \(N\). This tree has \(N\) leaf nodes, and these conventional decoders estimate the transmitted data bits by traversing the entire tree and examining the leaf nodes. However, if all the leaf nodes of an intermediate node in the polarization tree are frozen bits (known as a rate-0 node), there is no need to traverse the corresponding subtree, as the values of the frozen bits are already known to the decoder. The same thing happens when all the leaf nodes of an intermediate node are information bits (known as a rate-1 node). Pruning the polarization tree by eliminating the need to visit the sub-trees corresponding to rate-1 and rate-0 nodes can significantly reduce the complexity and latency of decoders \cite{alamdar2011simplified}.
This advantage becomes even more evident for low-rate (with many rate-0 nodes) or high-rate (with many rate-1 nodes) polar codes. The decoders resulting from pruning the polarization tree are referred to as fast SC (FSC) or fast SCL (FSCL) decoders. The channel polarization process can also be adjusted to account for rate-0 and rate-1 nodes \cite{el2017relaxed}. Also, fast SCL and Fano decoders, which are employed in the decoding of PAC codes \cite{zhu2023fast, zhao2023fast, ji2023low}, do not often visit the last level (and may visit only up to the one before last) of the polarization tree. 

The implementation of fast decoders necessitates calculating the metric function optimally at the intermediate levels of the polarization tree. In \cite{moradi2023tree}, the metric function used in the conventional SCL decoding algorithm is analyzed.
In \cite{zhang2015split}, assuming a Gaussian distribution, an ad-hoc method is proposed to avoid splitting on reliable bits by defining a threshold on the reliability of an information bit. 
In \cite{yao2024low}, the average of the log-likelihood ratios (LLRs) are empirically obtained, and based on that, they avoid splitting. 
They generalize the REP and Type-I nodes to any intermediate node with one or two information bits. 
This approach requires additional memory and computation to partially encode the related chunk of leaf nodes. 
In the rate profiles studied in this paper, we only see the information bit positions at the end of the chunk, although our method can be generalized in a straightforward fashion as well.
A recursive implementation of a fast SCL decoding was also explored in \cite{dumer2006soft}.

\subsection{Stack decoding}
Stack decoding is a sequential decoding algorithm that explores the code tree by maintaining all explored paths in a stack~\cite{zigangirov1966some, jelinek1969fast}.
The algorithm aims to identify the path on the code tree that corresponds to the transmitted codeword, starting at the root of the tree with an initial path metric of zero.
The decoder continues exploring the tree until it reaches a leaf node or satisfies a predefined stopping criterion.
At each iteration, the decoder selects (pops) the path with the highest metric from the stack. If the current bit is an information bit, the algorithm bifurcates the path, computes the metrics of both successor paths, and pushes them back into the stack. If the bit is frozen, only the single valid successor path is extended, and its metric is computed and stored.
For a memory-efficient implementation of stack decoding for PAC codes, we refer the reader to~\cite[Ch. 6]{moradi2022performance}.

\section{Metric Polarization}\label{sec: MetricPolarization}

Let \( W^N \) represents \( N \) independent and identically distributed (i.i.d.) copies of the channel \( W \). Let \( \mathbf{x}^N \) denote the length-$N$ input to \( W^N \) and \( \mathbf{y}^N \) denote the length-$N$ output in an uncoded system.
Consider a decoder that aims to estimate the inputs of channel \( W^N \) by using the channel output values \( \mathbf{y}^N \). Then the optimal approach, in terms of minimizing error probability, is to find the \( x_i \) values that maximize 
\begin{equation}
    p(x_i|y_i)
\end{equation}
known as the maximum a posteriori (MAP) rule. 
Using Bayes' rule, this can be expressed as
\begin{equation}
    p(x_i|y_i) = \frac{p(y_i|x_i)}{p(y_i)}p(x_i),
\end{equation}
where \( p(y_i|x_i) \) is the channel transition probability, \( p(y_i) \) is the channel output probability, and \( p(x_i) \) is the channel input probability. 

In the absence of prior knowledge about the input distribution, or when there is no coding, one may assume a uniform distribution for the \( x_i \) values. Thus, the MAP rule simplifies to maximizing the metric function
\begin{equation}\label{eq: RootMetric}
    \phi(x_i;y_i) \triangleq \log_2 \left( \frac{p(y_i|x_i)}{p(y_i)} \right).
\end{equation}
This metric function can be expressed in terms of the channel log-likelihood ratio (LLR) values as 
\begin{equation}
\begin{split}
    \phi(x_i;y_i) 
    &
    = \log_2 \left( \frac{p(y_i|x_i)}{p(y_i)} \right)\\ 
    & 
    = \log_2 \left( \frac{p(y_i|x_i)}{\frac{1}{2}\left[ p(y_i|x_i) + p(y_i|x_i \oplus 1) \right]} \right)\\
    &
    = 1 - \log_2\left(1 + \frac{p(y_i|x_i \oplus 1)}{p(y_i|x_i)} \right)\\
    &
    = 1 - \log_2\left(1 + 2^{-\log_2\left(\frac{p(y_i|x_i)}{p(y_i|x_i \oplus 1)}\right)}  \right)\\
    &
    = 1 - \log_2\left(1 + 2^{-L_{\text{ch}_i} \cdot (-1)^{x_i}}  \right),
\end{split}
\end{equation}
where 
\begin{equation}
    L_{\text{ch}_i} = \log_2 \frac{p(y_i|x_i = 0)}{p(y_i|x_i = 1)}
\end{equation}
and for a binary input additive white Gaussian noise (BI-AWGN) channel, this is equal to \( \frac{2y_i}{\sigma^2} \).

In this trivial decoding algorithm for the uncoded case, the metric function determines the direction (whether \( x_i \) is zero or one) to be chosen through the code tree. This metric function should have well-diverged values for the correct and incorrect values of \( x_i \). One way to analyze this problem is by considering the ensemble of input-output random pairs \((X_i, Y_i)\) and evaluating the average of the metric function. The expectation of the metric can be obtained as

\begin{equation}\label{eq: I(W)}
\begin{split}
    \mathbb{E}_{X_i,Y_i}\left[ \phi (X_i;Y_i)\right] 
    &
    = \sum_{x_i} p(x_i) \sum_{y_i} p(y_i|x_i) \phi(x_i;y_i) \\
    & 
    = \sum_{x_i} p(x_i) \sum_{y_i} p(y_i|x_i) \log_2\left(\frac{p(y_i|x_i)}{p(y_i)}\right)\\
    &
    = I(W),
\end{split}  
\end{equation}
where \( I(W) \) is the symmetric capacity, representing the highest achievable rate when the input \( x_i \) values have equal probabilities. Assuming that \( \tilde{x}_i = x_i \oplus 1 \), we have
\begin{equation}
\begin{split}
    \mathbb{E}_{X_i,Y_i}\left[ \phi (\tilde{X}_i;Y_i)\right] 
    &
    \leq 0.
\end{split}  
\end{equation}

In the \textit{high-capacity} regime, i.e., when the channel \( W \) is reliable, \(\mathbb{E}_{X_i,Y_i}\left[ \phi (X_i;Y_i)\right] = I(W) \) is close to 1. Given that the average of the metric function on the incorrect input bit is less than or equal to 0 (and in the perfect noiseless channel $\phi (\tilde{x}_i;y_i) = -\infty$), we conclude that in a relatively good channel, the soft value of this metric can effectively distinguish between the correct and incorrect values of the channel input.

\begin{figure}[htbp]
    \centering
    \begin{tikzpicture}[thick]

        \node at (0,0) (u1) {$\mathbf{s}_{1}^{N/2~~~}$};
        \node at (0,-1) (u2) {$\mathbf{s}_{N/2 +1}^{N}$};

        \node[draw, circle, inner sep=0pt, minimum size=3mm] at (1.5,0) (DotNo1) {};
        \node[draw, circle, fill=black, inner sep=0pt, minimum size=2pt] at (1.5,-1) (dot_u1) {};

        \draw (u1) -- (DotNo1.east);
        \draw (DotNo1.east) -- (3,0) node[midway, above] {$\mathbf{x}_{1}^{N/2~~~}$};
        \draw (u2) -- (dot_u1.west);
        \draw (dot_u1.east) -- (3,-1) node[midway, above] {$\mathbf{x}_{N/2 +1}^{N}$};

        \draw (DotNo1.north) -- (dot_u1);

        \node[draw, rectangle] at (4,0) (W1) {$W^{N/2}$};
        \node[draw, rectangle] at (4,-1) (W2) {$W^{N/2}$};
        
        \node at (5.5,0) (y1) {$\mathbf{y}_{1}^{N/2~~~}$};
        \node at (5.5,-1) (y2) {$\mathbf{y}_{N/2 +1}^{N}$};

        \draw (2,0) -- (W1.west);
        \draw (W1.east) -- (y1);
        \draw (2,-1) -- (W2.west);
        \draw (W2.east) -- (y2);

    \end{tikzpicture}
    \caption{Flowchart of a one-step polarization scheme.}
    \label{fig: OneStepPolarization}
\end{figure}

Fig. \ref{fig: OneStepPolarization} shows the one-step polarization process, which generates \(N/2\) i.i.d. bad channels named \(W^-\) and \(N/2\) i.i.d. good channels named \(W^+\) from \(N\) copies of an i.i.d. channel \(W\). The input to the \(i\)-th bad channel is \(s_i\), and its output is the pair \((y_i, y_{N/2 +i})\). The input to the \(i\)-th good channel is \(s_{N/2 +i}\), and assuming that a genie correctly provides the value of \(s_i\) for the \(i\)-th bad channel, the output of the \(i\)-th good channel is \((y_i, y_{N/2 +i}, s_i)\).

For the \(i\)-th bad channel, an optimal decoder should maximize 
\begin{equation}
    P(s_i|y_i,y_{N/2 +i}).
\end{equation}
Assuming a uniform distribution for its input \(s_i\), by using Bayes' rule, an optimal decoder should maximize the value of the metric function
\begin{equation}
    \phi^-(s_i; y_i,y_{N/2+i}) \triangleq \log_2 \left(\frac{P(y_i,y_{N/2+i}|s_i)}{P(y_i,y_{N/2+i})}\right).
\end{equation}
After a few steps of operation, we obtain
\begin{equation}
    \phi^-(s_i; y_i,y_{N/2+i}) = 1 - \log_2\left(1 + 2^{-L_i \cdot (-1)^{s_i}}\right),
\end{equation}
where
\begin{equation}
    L_i = \log_2\left(\frac{P(y_i, y_{N/2+i} \mid s_i = 0)}{P(y_i, y_{N/2+i} \mid s_i = 1)}\right).
\end{equation}

For the \(i\)-th good channel, similarly, an optimal decoder should maximize
\begin{equation}
\begin{split}
    &\phi^+(s_{N/2+i}; y_i,y_{N/2+i}, s_i) \\
    &~~~
    \triangleq \log_2 \left(\frac{P(y_i,y_{N/2+i}, s_i|s_{N/2+i})}{P(y_i,y_{N/2+i})}\right).
\end{split}
\end{equation}

By considering the ensemble of input-output random pairs \((S_i;Y_i,Y_{N/2 +i})\) for the \(i\)-th bad channel, the average of the corresponding metric is
\begin{equation}
\begin{split}
    & \mathbb{E}_{S_i,(Y_i,Y_{N/2+i})}\left[\phi^-(S_i; Y_i,Y_{N/2+i})\right] \\
    & = \sum_{s_i} p(s_i) \sum_{(y_i,y_{N/2+i})} P(y_i,y_{N/2+i}|s_i) \phi^-(s_i; y_i,y_{N/2+i}) \\
    & = \sum_{s_i} \sum_{(Y_i,Y_{N/2+i})} p(s_i) P(y_i,y_{N/2+i}|s_i) \\
    &~~~~~~~~~~~~~~~~~~~~~\times \log_2 \left(\frac{P(y_i,y_{N/2+i}|s_i)}{P(y_i,y_{N/2+i})}\right) = I(W^-).
\end{split}
\end{equation}

Assuming that \(\tilde{s}_i = s_i \oplus 1\), we can also show that
\begin{equation}
    \mathbb{E}_{S_i,(Y_i,Y_{N/2+i})}\left[\phi^-(\tilde{S}_i; Y_i,Y_{N/2+i})\right] \leq 0.
\end{equation}

Similarly, for the \(i\)-th good channel \(W^+\), the average of its metric function is
\begin{equation}
    \mathbb{E}\left[ \phi^+(S_{N/2+i}; Y_i,Y_{N/2+i}, S_i) \right] = I(W^+),
\end{equation}
and for \(\tilde{s}_{N/2+i} = s_{N/2+i} \oplus 1\),
\begin{equation}
    \mathbb{E}\left[ \phi^+(\tilde{S}_{N/2+i}; Y_i,Y_{N/2+i}, S_i) \right] \leq 0.
\end{equation}

\begin{figure}[htbp] 
\centering
\resizebox{\columnwidth}{!}{
\begin{tikzpicture}[thick]

    \node at (0,0) (u1) {$\mathbf{v}_1^{N/4~~~~}$};
    \node at (0,-1) (u2) {$\mathbf{v}_{N/4+1~}^{N/2}$};
    \node at (0,-2) (u3) {$\mathbf{v}_{N/2+1~}^{3N/4}$};
    \node at (0,-3) (u4) {$\mathbf{v}_{3N/4+1}^{N}$};

    \node[draw, circle, inner sep=0pt, minimum size=3mm] at (1.5,0) (DotNo1) {};
    \node[draw, circle, fill=black, inner sep=0pt, minimum size=2pt] at (1.5,-1) (dot_u1) {};
    \node[draw, circle, inner sep=0pt, minimum size=3mm] at (1.5,-2) (DotNo2) {};
    \node[draw, circle, fill=black, inner sep=0pt, minimum size=2pt] at (1.5,-3) (dot_u2) {};

    \draw (u1) -- (DotNo1.east);
    \draw (DotNo1.east) -- (2.75,0) node[midway, above] {$\mathbf{s}_1^{N/4~~~~}$} -- (4,0) -- (5,0) node[midway, above] {$\mathbf{x}_1^{N/4~~~~}$};
    \draw (u2) -- (dot_u1.west);
    \draw (dot_u1.east) -- (2.75,-1) node[midway, above] {$\mathbf{s}_{N/4+1~}^{N/2}$} -- (4,-1) -- (5,-1) node[midway, above] {$\mathbf{x}_{N/4+1~}^{N/2}$};
    \draw (dot_u1) -- (DotNo1.north);
    \draw (u3) -- (DotNo2.east);
    \draw (DotNo2.east) -- (2.75,-2) node[midway, above] {$\mathbf{s}_{N/2+1~}^{3N/4}$} -- (4,-2) -- (5,-2) node[midway, above] {$\mathbf{x}_{N/2+1~}^{3N/4}$};
    \draw (u4) -- (dot_u2.west);
    \draw (dot_u2.east) -- (2.75,-3) node[midway, above] {$\mathbf{s}_{3N/4+1}^{N}$} -- (4,-3) -- (5,-3) node[midway, above] {$\mathbf{x}_{3N/4+1}^{N}$};
    \draw (dot_u2) -- (DotNo2.north);

    \node[draw, circle, inner sep=0pt, minimum size=3mm] at (3,0) (DotNo3) {};
    \node[draw, circle, inner sep=0pt, minimum size=3mm] at (3.5,-1) (DotNo4) {};
    \node[draw, circle, fill=black, inner sep=0pt, minimum size=2pt] at (3,-2) (dot_v1) {};
    \node[draw, circle, fill=black, inner sep=0pt, minimum size=2pt] at (3.5,-3) (dot_v2) {};

    \draw (DotNo3) -- (4,0);
    \draw (DotNo4) -- (4,-1);
    \draw (DotNo3.north) -- (dot_v1);
    \draw (DotNo4.north) -- (dot_v2);

    \node[draw, rectangle] at (6,0) (W1) {$W^{N/4}$};
    \node[draw, rectangle] at (6,-1) (W2) {$W^{N/4}$};
    \node[draw, rectangle] at (6,-2) (W3) {$W^{N/4}$};
    \node[draw, rectangle] at (6,-3) (W4) {$W^{N/4}$};

    \node at (8,0) (y1) {$\mathbf{y}_1^{N/4~~~}$};
    \node at (8,-1) (y2) {$\mathbf{y}_{N/4+1~}^{N/2}$};
    \node at (8,-2) (y3) {$\mathbf{y}_{N/2+1~}^{3N/4}$};
    \node at (8,-3) (y4) {$\mathbf{y}_{3N/4+1}^{N}$};

    \draw (4,0) -- (W1.west);
    \draw (W1.east) -- (y1);
    \draw (4,-1) -- (W2.west);
    \draw (W2.east) -- (y2);
    \draw (4,-2) -- (W3.west);
    \draw (W3.east) -- (y3);
    \draw (4,-3) -- (W4.west);
    \draw (W4.east) -- (y4);

\end{tikzpicture}
}
    \caption{Flowchart of a two-step polarization scheme.} 
    \label{fig: TwoStepPolarization}
\end{figure}


Continuing with the recursive process, Fig. \ref{fig: TwoStepPolarization} illustrates the two-step polarization. The metric function \(\phi^{--}\) corresponding to channel \(W^{--}\) is defined as
\begin{equation}
\begin{split}
    & \phi^{--}(v_i;y_i,y_{N/4+i},y_{N/2+i},y_{3N/4+i}) \\
    & \triangleq \log_2 \left( \frac{P(y_i,y_{N/4+i},y_{N/2+i},y_{3N/4+i}|v_i)}{P(y_i,y_{N/4+i},y_{N/2+i},y_{3N/4+i})} \right).
\end{split}
\end{equation}
Similarly, for channels \(W^{-+}\), \(W^{+-}\), and \(W^{++}\), we define the metric functions \(\phi^{-+}\), \(\phi^{+-}\), and \(\phi^{++}\). Assuming the inputs of the synthesized channels are uniformly distributed, an optimal decoder should maximize the value of these metric functions.

In the same way, we can show that these optimal metric functions are equal to the corresponding channel capacities of the four synthesized channels when considering the correct input bit, and they have a negative value when the input bit is incorrect.

Continuing this process provides the polarization of the metric function on the polarization tree and is obtained by the LLRs at the corresponding depth of the tree. 
Assuming the input to the synthesized channel is uniform, maximizing the polarized metric functions results in an optimal decoder that can partially explore the polarization tree, similar to FSCL decoders.

\begin{figure}[h]
    \centering
    \begin{tabular}{cc}
    \adjustbox{valign=c}{\subfloat[\label{fig: PolarizationTree a}]{%
          \begin{tikzpicture}
              [level distance=15mm,
               every node/.style={circle, draw, fill, inner sep=1.5pt},
               level 1/.style={sibling distance=20mm},
               level 2/.style={sibling distance=10mm},
               edge from parent/.style={draw, thick}]
              \node (root) {}
                 child {node {}
                   child {node {}}
                   child {node {}}
                 }
                 child {node {}
                   child {node {}}
                   child {node {}}
                 };
            
              \node at (-1.5, -3.3) [circle, draw=none, fill=none, inner sep=1pt, minimum size=5mm] {$\phi^{--}$};
              \node at (-.5, -3.3) [circle, draw=none, fill=none, inner sep=1pt, minimum size=5mm] {$\phi^{-+}$};
              \node at (.5, -3.3) [circle, draw=none, fill=none, inner sep=1pt, minimum size=5mm] {$\phi^{+-}$};
              \node at (1.5, -3.3) [circle, draw=none, fill=none, inner sep=1pt, minimum size=5mm] {$\phi^{++}$};
            
              \draw[->, thick] (-.25, 0) -- (-1, 0) node[draw=none, fill=none, midway, above] {$-$};
              \draw[->, thick] (0.25, 0) -- (1, 0) node[draw=none, fill=none, midway, above] {$+$};
            
            \end{tikzpicture}
          
          }          
          }
    &      
    \adjustbox{valign=c}{\subfloat[\label{fig: CodeTree b}]{%
          \begin{tikzpicture}
            [grow=east, sibling distance=2.5cm, level distance=0.75cm, 
                every node/.style={circle, draw, fill, inner sep=1.5pt},
                level 1/.style={sibling distance=0mm},
                level 2/.style={sibling distance=20mm},
                level 3/.style={sibling distance=10mm},
                level 4/.style={sibling distance=0mm},
                edge from parent/.style={draw, thick}]
            
                \node {}
                child {node {}
                    child {node {}
                        child {node {} child{node {}}}
                        child {node {} child{node {}}}
                    }
                    child {node {}
                        child {node {} child{node {}}}
                        child {node {} child{node {}}}
                    }
                };

                \draw[->, thick] (0, -.25) -- (0, -1) node[draw=none, fill=none, midway, right] {$1$};
                \draw[->, thick] (0, 0.25) -- (0, 1) node[draw=none, fill=none, midway, right] {$0$};
            
            \end{tikzpicture}
          
          }
          }
    \end{tabular}
    \caption{(a) Polarization and (b) code tree of a PAC$(4,2)$ code.}
    \label{fig: CodeTree} 
\end{figure}

Fig. \ref{fig: CodeTree} illustrates the polarization tree and code tree corresponding to a PAC$(4,2)$ code, where $\mathcal{A} = \{2,3\}$. For $K$ data bits, the code tree contains $2^K$ distinct paths from the root to each of the leaf nodes. The decoder's task is to identify the path that corresponds to the transmitted data. This involves determining whether to proceed up (corresponding to data bit $0$) or down (corresponding to data bit $1$).
The polarization tree has at most $\log_2(N)$ levels. 
Using the polarization tree, the decoder computes the bit metric function values and subsequently traverses the code tree. Each leaf node of the polarization tree corresponds to a specific level of the code tree. For a Fast SCL decoder, it is not necessary to traverse the polarization tree up to the leaf level to determine the correct branch in the code tree.
The following proposition characterizes the polarization of the metric function.
As discussed in \cite[p. 206]{gallager1968information}, we employ the standard ensemble of random linear codes for convolutional codes to conduct our analysis. This ensemble can be represented by the pair $(\mathbf{T}, \mathbf{c})$ in the form $ \mathbf{v}\mathbf{T} + \mathbf{c}$, where $\mathbf{c}$ is an arbitrary, fixed binary vector of length $N$.

We can now present the general result.
Consider a convolutional code with output $\mathbf{r} = \mathbf{v}\mathbf{T} + \mathbf{c}$, where the polar mapper undergoes $k$ steps of polarization. Define
\begin{equation*}
    \prescript{}{i}{\mathbf{r}} \triangleq \left(r_i, r_{N/2^k + i}, \ldots, r_{(2^k-1)N/2^k + i}\right)
\end{equation*}
and its corresponding channel output as 
\begin{equation*}
     \prescript{}{i}{\mathbf{y}} \triangleq \left(y_i, y_{N/2^k + i}, \ldots, y_{(2^k-1)N/2^k + i}\right)
\end{equation*}
for $i$ ranging from 1 to $N/2^k$.
By employing $k$ steps of polarization, we have $2^k$ of $N/2^k$ i.i.d. channels 
$W^{\{-,+\}^{k}}(\prescript{}{i}{r}_j; \prescript{}{i}{\mathbf{y}}, \prescript{}{i}{\mathbf{r}}_1^{j-1}) $ 
from the original $N$ i.i.d. channels $W$. 
Assume the decoder has obtained $\prescript{}{i}{\mathbf{r}}_1^{j-1}$. 

\begin{proposition} \label{proposition}
The optimal metric function for decoding $\prescript{}{i}{r}_j$ is 
\begin{equation}
\phi^{\{-,+\}^{k}}(\prescript{}{i}{r}_j; 
     \prescript{}{i}{\mathbf{y}}, \prescript{}{i}{\mathbf{r}}_1^{j-1}) 
    \triangleq \log_2 \left( \frac{P(\prescript{}{i}{\mathbf{y}}, \prescript{}{i}{\mathbf{r}}_1^{j-1} \mid \prescript{}{i}{r}_j)}{P(\prescript{}{i}{\mathbf{y}}, \prescript{}{i}{\mathbf{r}}_1^{j-1})} \right),
\end{equation}
where $i$ ranges from 1 to $N/2^k$, corresponding to the $i$-th channel of $W^{\{-,+\}^{k}}(\prescript{}{i}{r}_j; 
     \prescript{}{i}{\mathbf{y}}, \prescript{}{i}{\mathbf{r}}_1^{j-1}) $. The average value of this metric function for the correct branch is $I\left(W^{\{-,+\}^{k}}(\prescript{}{i}{r}_j; 
     \prescript{}{i}{\mathbf{y}}, \prescript{}{i}{\mathbf{r}}_1^{j-1}) \right)$, while for the incorrect branch, it is less than or equal to zero.
\end{proposition}


\textit{Proof:}
By the MAP rule, an optimal decoder seeks to maximize 
$P(\prescript{}{i}{r}_j \mid \prescript{}{i}{\mathbf{y}}, \prescript{}{i}{\mathbf{r}}_1^{j-1})$.
Utilizing Bayes' rule, this can be expressed as
\begin{equation*}
    P(\prescript{}{i}{r}_j \mid \prescript{}{i}{\mathbf{y}}, \prescript{}{i}{\mathbf{r}}_1^{j-1}) =
     \frac{P(\prescript{}{i}{\mathbf{y}}, \prescript{}{i}{\mathbf{r}}_1^{j-1} \mid \prescript{}{i}{r}_j)}{P(\prescript{}{i}{\mathbf{y}}, \prescript{}{i}{\mathbf{r}}_1^{j-1})}
     P(\prescript{}{i}{r}_j).
\end{equation*}
Applying the monotonically increasing $\log_2$ function and assuming uniform distribution on the input of the synthesized channel, the MAP rule leads to maximizing
\begin{equation}
    \phi^{\{-,+\}^{k}}(\prescript{}{i}{r}_j; 
     \prescript{}{i}{\mathbf{y}}, \prescript{}{i}{\mathbf{r}}_1^{j-1}) 
    \triangleq
    \log_2 \left( \frac{P(\prescript{}{i}{\mathbf{y}}, \prescript{}{i}{\mathbf{r}}_1^{j-1} \mid \prescript{}{i}{r}_j)}{P(\prescript{}{i}{\mathbf{y}}, \prescript{}{i}{\mathbf{r}}_1^{j-1})} \right).
\end{equation}
Considering the ensemble of input-output random pairs 
$(\prescript{}{i}{R}_j; \prescript{}{i}{\mathbf{Y}}, \prescript{}{i}{\mathbf{R}}_1^{j-1})$, the average value of the corresponding metric is
\begin{equation}
\begin{split}
    & \mathbb{E}_{
    \prescript{}{i}{R}_j,
    ( \prescript{}{i}{\mathbf{Y}}, \prescript{}{i}{\mathbf{R}}_1^{j-1})
     } 
     \left[ 
    \phi^{\{-,+\}^{k}}(\prescript{}{i}{R}_j; 
     \prescript{}{i}{\mathbf{Y}}, \prescript{}{i}{\mathbf{R}}_1^{j-1})  
     \right]\\
     & =
     \sum_{\prescript{}{i}{r}_j}
     P(\prescript{}{i}{r}_j)
     \sum_{(\prescript{}{i}{\mathbf{y}}, \prescript{}{i}{\mathbf{r}}_1^{j-1})}
     P(\prescript{}{i}{\mathbf{y}}, \prescript{}{i}{\mathbf{r}}_1^{j-1} \mid \prescript{}{i}{r}_j)  \\
     & ~~~~~~~~~~~\times
     \phi^{\{-,+\}^{k}}(\prescript{}{i}{r}_j; 
     \prescript{}{i}{\mathbf{y}}, \prescript{}{i}{\mathbf{r}}_1^{j-1}) \\
     & =
     \sum_{\prescript{}{i}{r}_j}
     \sum_{(\prescript{}{i}{\mathbf{y}}, \prescript{}{i}{\mathbf{r}}_1^{j-1})}
     P(\prescript{}{i}{r}_j)
     P(\prescript{}{i}{\mathbf{y}}, \prescript{}{i}{\mathbf{r}}_1^{j-1} \mid \prescript{}{i}{r}_j)  \\
     & ~~~~~~~~~~~\times
     \log_2 \left( \frac{P(\prescript{}{i}{\mathbf{y}}, \prescript{}{i}{\mathbf{r}}_1^{j-1} \mid \prescript{}{i}{r}_j)}{P(\prescript{}{i}{\mathbf{y}}, \prescript{}{i}{\mathbf{r}}_1^{j-1})} \right)\\
     & = I\left(W^{\{-,+\}^{k}}(\prescript{}{i}{r}_j; 
     \prescript{}{i}{\mathbf{y}}, \prescript{}{i}{\mathbf{r}}_1^{j-1}) \right). \\
\end{split}
\end{equation}
Assuming that $\Tilde{\prescript{}{i}{r}_j} = \prescript{}{i}{r}_j \oplus 1$, we have
\begin{equation}
\begin{split}
    & \mathbb{E}_{
    \prescript{}{i}{R}_j,
    \Tilde{\prescript{}{i}{R}_j},
    ( \prescript{}{i}{\mathbf{Y}}, \prescript{}{i}{\mathbf{R}}_1^{j-1})
     } 
     \left[ 
    \phi^{\{-,+\}^{k}}(\Tilde{\prescript{}{i}{R}_j}; 
     \prescript{}{i}{\mathbf{Y}}, \prescript{}{i}{\mathbf{R}}_1^{j-1})  
     \right]  \\
     & =
     \sum_{\Tilde{\prescript{}{i}{r}_j}} q(\Tilde{\prescript{}{i}{r}_j})
     \sum_{\prescript{}{i}{r}_j} q(\prescript{}{i}{r}_j)
     \sum_{(\prescript{}{i}{\mathbf{y}}, \prescript{}{i}{\mathbf{r}}_1^{j-1})}
     P(\prescript{}{i}{\mathbf{y}}, \prescript{}{i}{\mathbf{r}}_1^{j-1} \mid \prescript{}{i}{r}_j)\\
     & ~~~~~~~~~~~~~~~~~~~~~~~~~~~~~~\times
     \phi^{\{-,+\}^{k}}(\Tilde{\prescript{}{i}{R}_j}; 
     \prescript{}{i}{\mathbf{Y}}, \prescript{}{i}{\mathbf{R}}_1^{j-1})  \\
     & =
     \sum_{\Tilde{\prescript{}{i}{r}_j}} 
     \sum_{(\prescript{}{i}{\mathbf{y}}, \prescript{}{i}{\mathbf{r}}_1^{j-1})}
     q(\Tilde{\prescript{}{i}{r}_j})
     P(\prescript{}{i}{\mathbf{y}}, \prescript{}{i}{\mathbf{r}}_1^{j-1} )
     \log_2 \left( \frac{P(\prescript{}{i}{\mathbf{y}}, \prescript{}{i}{\mathbf{r}}_1^{j-1} \mid \Tilde{\prescript{}{i}{r}_j})}{P(\prescript{}{i}{\mathbf{y}}, \prescript{}{i}{\mathbf{r}}_1^{j-1})} \right)\\
     & \leq\frac{1}{\ln(2)} 
     \sum_{\Tilde{\prescript{}{i}{r}_j}} 
     \sum_{(\prescript{}{i}{\mathbf{y}}, \prescript{}{i}{\mathbf{r}}_1^{j-1})}
     q(\Tilde{\prescript{}{i}{r}_j})
     P(\prescript{}{i}{\mathbf{y}}, \prescript{}{i}{\mathbf{r}}_1^{j-1} )\\
     &~~~~~~~~~~~~~~~~~~~~~~~~~~~~~~\times
     \left[ \frac{P(\prescript{}{i}{\mathbf{y}}, \prescript{}{i}{\mathbf{r}}_1^{j-1} \mid \Tilde{\prescript{}{i}{r}_j})}{P(\prescript{}{i}{\mathbf{y}}, \prescript{}{i}{\mathbf{r}}_1^{j-1})} - 1\right] = 0,
\end{split}
\end{equation}
where we used $\ln(x) \leq x-1$ for $x>0.$
\hfill\IEEEQEDhere

In addition to the mean, analyzing the variance of the polarized metric function values is essential because it reveals how spread out these values are around the polarized mutual information. If the variance is large, the polarized mutual information, as the mean, can give a misleading impression of the \textit{typical} polarized metric function value. We show that for general channels, the variances of the bit metric values approach zero, allowing the polarized mutual information to be considered the typical value.

The variance of the metric can also be obtained as
\begin{equation}\label{eq: V(W)}
\begin{split}
    & \text{Var}_{X_i,Y_i}\left[ \phi (X_i;Y_i)\right] \\
    &~~~
    = \sum_{x_i} p(x_i) \sum_{y_i} p(y_i|x_i) \phi^{2}(x_i;y_i) - I(W)^{2} \\
    &~~~
    = \sum_{x_i} p(x_i) \sum_{y_i} p(y_i|x_i) \log_2^{2}\left(\frac{p(y_i|x_i)}{p(y_i)}\right)  - I(W)^{2} \\
    &~~~
    = V(W),
\end{split}  
\end{equation}
where \( V(W) \) is the channel dispersion.
In a finite-blocklength regime, the maximum coding rate can be obtained from $I(W)$ and $V(W)$ \cite{polyanskiy2010channel} and \eqref{eq: I(W)} and \eqref{eq: V(W)} relate these parameters to the SCL decoding metric function.

For a BI-AWGN channel with BPSK modulation and noise variance $\sigma^2$, it is shown in \cite{brannstrom2004convergence} that $I(W) = J(t)$, where
\begin{equation}
    J(t) \triangleq 1 - \frac{1}{t\sqrt{2\pi}} \int_{-\infty}^{+\infty} e^{\frac{-(u - \frac{t^2}{2})^2}{2t^2}} \log_2(1 + e^{-u}) \, du,
\end{equation}
with $t = \frac{2}{\sigma}$.
An approximation for \( J(t) \) is also provided by
\begin{equation}
    J_{\text{approx}}(t) = \left[1 - 2^{-0.3073 \, t^{2 \times 0.8935}} \right]^{1.1064}.
\end{equation}

For this BI-AWGN channel with BPSK modulation, the variance of the metric function variable can be obtained as
\begin{equation}
\begin{split}
    & \text{Var}_{X,Y}\left[\phi(X,Y)\right]\\
    &= \sum_{x\in\{-1,1\}}\int_{-\infty}^{+\infty}\frac{1}{2}p(y|x)\phi^2(X,Y) dy - I(W)^2.
\end{split}
\end{equation}
By performing some calculus, we derive
\begin{equation}
    \text{Var}_{X,Y}\left[\phi(X,Y)\right]  = -(J(t) - 1)^2 - K(t) +1,
\end{equation}
where
\begin{equation}\label{eq:Kt}
    K(t) \triangleq 1 - \frac{1}{t \sqrt{2\pi}} \int_{-\infty}^{+\infty} 
e^{-\frac{(u - \frac{t^2}{2})^2}{2 t^2}} 
\log_2^2 \left( 1 + e^{-u} \right) dy.
\end{equation}
Similarly, using the Nelder-Mead simplex method, we derive an approximation for \( K(t) \) as
\begin{equation}\label{eq:Kt_approx}
    K_{\text{approx}}(t) = \left[1 - 2^{-0.96483 t^{2 \times 0.61746}} \right]^{10.232}.
\end{equation}

\begin{figure}[t] 
\centering
	\includegraphics [width = \columnwidth]{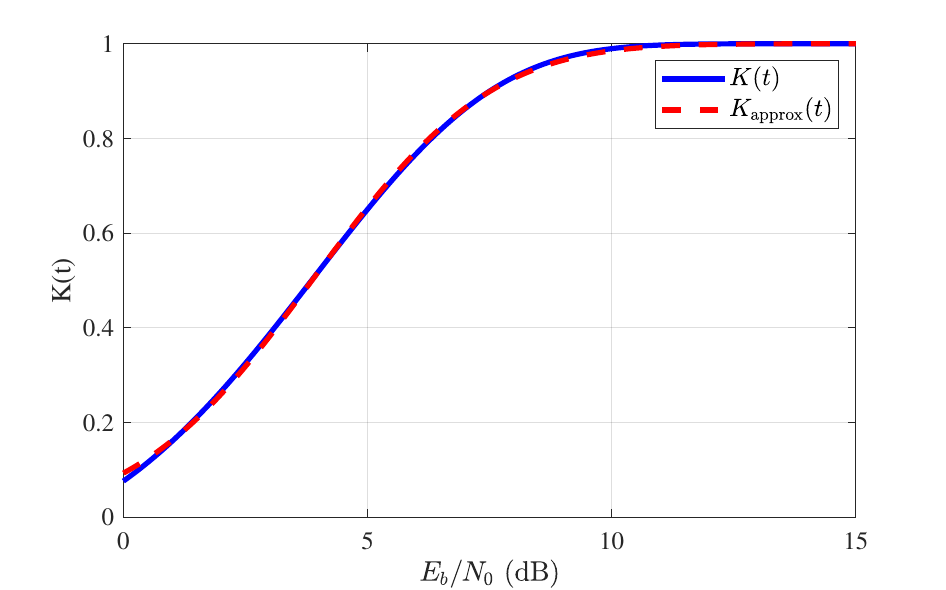}
	\caption{Comparison of $K(t)$ and its approximation $K_{\text{approx}}(t)$ for different values of $E_b/N_0$.} 
	\label{fig: AWGN_Kt_Compare}
\end{figure}

\begin{figure}[t] 
\centering
	\includegraphics [width = \columnwidth]{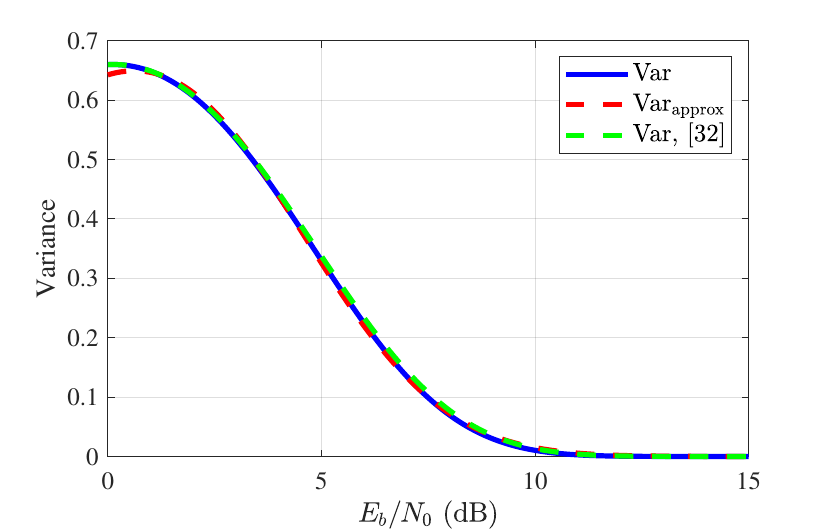}
	\caption{Comparison of $\text{Var}_{X,Y}\left[\phi(X,Y)\right]$ and its approximation using $J_{\text{approx}}(t)$ and $K_{\text{approx}}(t)$, along with the variance of the channel approximation obtained from the method in \cite{tal2013construct}.}
	\label{fig: AWGN_Var_Compare}
\end{figure}

\begin{figure}[t] 
\centering
	\includegraphics [width = \columnwidth]{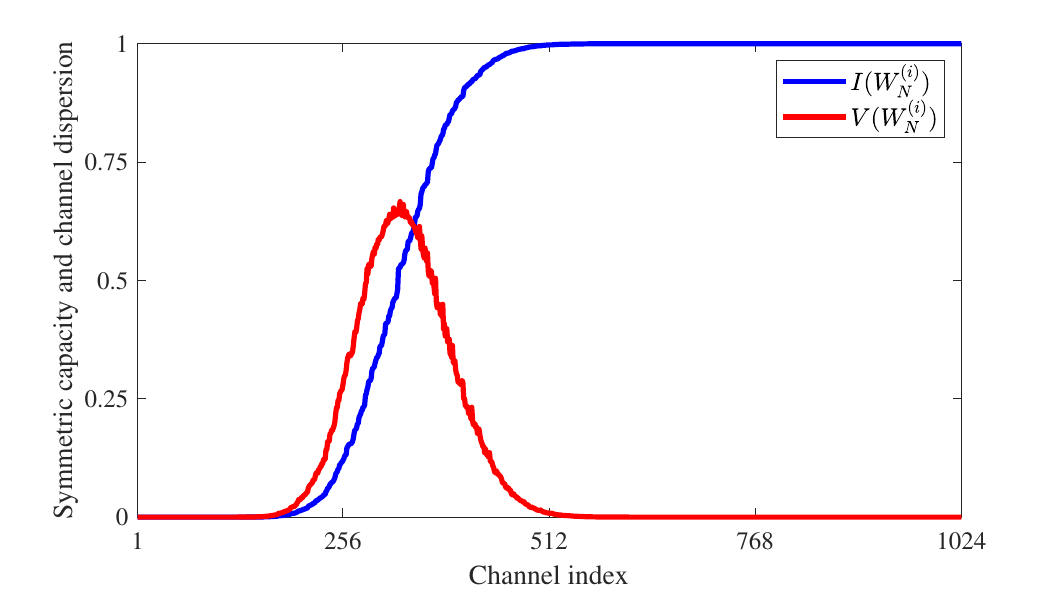}
	\caption{Plot of sorted $I(W_N^{(i)})$ and corresponding $V(W_N^{(i)})$ versus channel indices $1,2,\cdots, N = 1024$ for a BI-AWGN at $E_b/N_0=2.5$~dB.}
	\label{fig: AWGN_IW_VW_N1024}
\end{figure}

Fig.~\ref{fig: AWGN_Kt_Compare} plots $K(t)$ from \eqref{eq:Kt} and compares it with our proposed approximation in \eqref{eq:Kt_approx}.
Fig.~\ref{fig: AWGN_Var_Compare} also plots $\text{Var}_{X,Y}\left[\phi(X,Y)\right]$ using $K(t)$ and $J(t)$ and compares it with the variance approximation obtained from $J_{\text{approx}}(t)$ and $K_{\text{approx}}(t)$ functions. 
Additionally, we plot the variance of the channel approximation obtained using the method introduced in \cite{tal2013construct} for computing channel capacity.
As shown in this figure, to keep the variance of the metric values close to zero, $E_b/N_0$ should be greater than $10~$dB.  
At $2.5~$dB, the variance exceeds $0.5$.  
In theorem \ref{thm:Var}, we prove that for polar and PAC codes, the variance of the bit-metric values almost surely approaches zero.
For \( N = 1024 \), Fig. \ref{fig: AWGN_IW_VW_N1024} shows the sorted polarized bit-channel capacities and the corresponding variances of a BI-AWGN channel at $E_b/N_0 = 2.5$~dB.

\begin{figure}[t] 
\centering
	\includegraphics [width = \columnwidth]{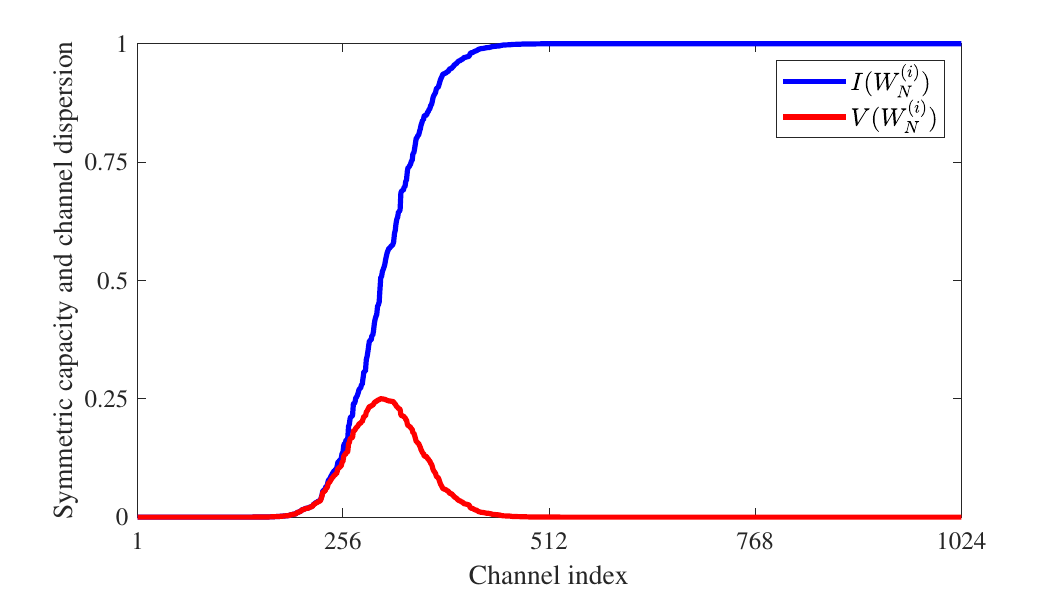}
	\caption{Plot of sorted $I(W_N^{(i)})$ and corresponding $V(W_N^{(i)})$ versus channel indices $1,2,\cdots, N = 1024$ for a BEC with $\epsilon=0.3$.} 
	\label{fig: BEC_IW_VW_N1024}
\end{figure}

For the case of a BEC($\epsilon$) with the erasure probability of $\epsilon$, the variance is $V = \epsilon(1-\epsilon)$.
For $\epsilon = 0.3$, the metric variance is $V = 0.21$.
Note that the maximum value of \( V \) achieves at \( \epsilon = 0.5 \), where it attains \( V = 0.25 \).
For \( N = 1024 \), Fig. \ref{fig: BEC_IW_VW_N1024} shows the sorted polarized bit-channel capacities and the corresponding variances of BEC$(0.3)$.
The following theorem proves the polarization effect for the variance of the polarized metric function values of BECs, and we later generalize this to any B-DMC.
\begin{theorem}
    For a BEC, the variances of the bit metrics approach zero.
\end{theorem}

\textit{Proof:}
For a BEC($\epsilon$), the mutual information is given by \(I(W) = 1 - \epsilon\). 
Furthermore, \(I(W_N^{(i)})\) converges a.e. to a random variable that takes values a.e. in \(\{0,1\}\).
Additionally, after each polarization step, a BEC remains a BEC. 
Consequently, \(V(W_N^{(i)})\) also converges a.e. to a random variable that takes a zero value a.e.

\hfill\IEEEQEDhere

For a binary symmetric channel (BSC) with crossover probability \(\delta\), the channel capacity \(I(W)\) is given by \(1 - h_2(\delta)\), where \(h_2(\delta)\) denotes the binary entropy function. 
The variance of the metric is given by
\[
V = \delta(1-\delta) \log_2^2\left(\frac{1-\delta}{\delta}\right).
\]
If the crossover probability \(\delta\) approaches \(0.5\) (i.e., when the channel is highly noisy), the capacity tends to zero. On the other hand, as \(\delta\) approaches \(0\) or \(1\) (i.e., for a noiseless channel), the capacity approaches 1. 
In each of these limiting cases, the channel dispersion \(V\) tends to zero.

The variance of the conditional entropy random variable, referred to as the varentropy of the channel, is a well-studied parameter that was originally explored by Strassen \cite{strassen1962asymptotische} to estimate the performance of finite blocklength coding. 
Ar{\i}kan proved that for a BI-DMC channel, the varentropies of polarized bit-channels approach zero asymptotically \cite{arikan2016varentropy}.
In the following theorem, we connect the polarization of the varentropy to the variances of the polarized metric functions.
\begin{theorem}\label{thm:Var}
For a BI-DMC, the variances of the bit metrics approach zero a.e.
\end{theorem}

\textit{Proof:}

For a BI-DMC with a uniform input and arbitrary output alphabet \( \mathcal{Y} \), denoted as \( W: \mathcal{X} \longrightarrow \mathcal{Y} \), the conditional entropy is related to the decoding metric function as
\begin{equation}
\begin{split}
    h(x|y) & \triangleq -\log_2 p(x|y) = -\log_2 \left( \frac{p(y|x)}{p(y)} \right) - \log_2 p(x) \\
    &= -\phi(x, y) + 1 = \log_2 \left( 1 + 2^{-L_{\text{ch}_i} \cdot (-1)^{x_i}} \right).
\end{split}
\end{equation}
Also,
\begin{equation}
    \text{Var}(\phi(X,Y)) = \text{Var}(1- h(X|Y)) = \text{Var}(h(X|Y)).
\end{equation}
Similarly, after $n$ steps of polarization,
\begin{equation}
\begin{split}
    h(u_i|\mathbf{y},\mathbf{u}^{i-1}) 
    &= -\log_2 P(u_i|\mathbf{y},\mathbf{u}^{i-1})\\
    &=- \log_2 \frac{P(\mathbf{y},\mathbf{u}^{i-1}|u_i)}{P(\mathbf{y},\mathbf{u}^{i-1})} - \log_2p(u_i) \\
    &= -\phi(u_i;\mathbf{y},\mathbf{u}^{i-1}) + 1,
\end{split}
\end{equation}
and
\begin{equation}
\begin{split}
    \text{Var}(\phi(U_i;\mathbf{Y},\mathbf{U}^{i-1})) 
    &= \text{Var}(1 - h(U_i|\mathbf{Y},\mathbf{U}^{i-1}) )\\
    &= \text{Var}(h(U_i|\mathbf{Y},\mathbf{U}^{i-1}) )
\end{split}
\end{equation}
For such a general channel, the varentropy also asymptotically decreases to zero a.e. \cite{arikan2016varentropy}.
Consequently, for the bit-channel \( W_N^{(i)} \), the variance of the corresponding bit-metric function random variable \( \phi(U_i; \mathbf{Y}, \mathbf{U}^{i-1}) \) also asymptotically decreases to zero a.e.

\hfill\IEEEQEDhere


The next theorem proves that the bit-metric value of a correct path concentrates around its bit-channel capacity for both almost noiseless and almost completely noisy channels. Since the list size of an SCL decoding algorithm and the search size of a sequential decoding algorithm depend on distinguishing the correct path from incorrect ones, this theorem suggests that the bit-channel variance can also be a crucial factor in determining the code rate profile, influencing the search complexity.

\begin{theorem}\label{thm:UpperBound}
For a BI-DMC, the $i$-th bit-metric random variable $\phi(U_i; \mathbf{Y}, \mathbf{U}^{i-1})$ converges to its mean in probability.
Specifically we have
\begin{equation}
P\left( \left| \phi(U_i; \mathbf{Y}, \mathbf{U}^{i-1}) - I(W_N^{(i)}) \right| \geq m_i \right)
 \leq \frac{\text{Var}(W_N^{(i)})}{m_i^2},
\end{equation}
\end{theorem}

\textit{Proof:}

With the mean and variance of the bit-metric function values in hand, Chebyshev's inequality states that
\begin{equation}
P\left( \left| \phi(U_i; \mathbf{Y}, \mathbf{U}^{i-1}) - I(W_N^{(i)}) \right| \geq m_i \right)
 \leq \frac{\text{Var}(W_N^{(i)})}{m_i^2},
\end{equation}
where \( m_i > 0 \) represents the deviation threshold from the expected value \( I(W_N^{(i)}) \). 
The previous theorem states that \( \text{Var}(W_N^{(i)}) \) approaches zero a.e.

\hfill\IEEEQEDhere


Based on Theorem \ref{thm:UpperBound}, we propose a list-pruning strategy for the SCL decoding algorithm. When decoding the $i$-th bit, if the probability that the distance of a bit-metric value from its mean exceeds \( m_i \) is greater than a threshold \( P_{th} \), we prune the corresponding decoding path.
Since \( 0 \leq I(W_N^{(i)}) \leq 1 \), we simply prune the paths for which the bit-metric value \( \phi(u_i; \mathbf{y}, \mathbf{u}^{i-1}) \) is less than \( m_i = -\sqrt{\frac{\text{Var}(W_N^{(i)})}{P_{th}}} \).
For a list size of $L$, if the number of pruned paths exceeds $L$ for the $i$-th bit, the sorting operation is no longer necessary.

\begin{figure*}[ht]
\centering
\begin{tikzpicture}
  [level distance=15mm,
   every node/.style={circle,inner sep=1pt},
   level 1/.style={sibling distance=90mm},
   level 2/.style={sibling distance=45mm},
   level 3/.style={sibling distance=15mm},
   level 4/.style={sibling distance=15mm}]
  \node {$0.7944$}
     child {node {$0.6386$}
       child {node {$0.4198$}
         child {node [rectangle split, rectangle split parts=2, text centered, draw=none] {0.1891\nodepart{second}\textbf{Rate-0}}}
         child {node {$0.6504$}
           child {node [rectangle split, rectangle split parts=2, text centered, draw=none] {0.4349\nodepart{second}\textbf{Rate-0}}}
           child {node [rectangle split, rectangle split parts=2, text centered, draw=none] {0.8663\nodepart{second}\textbf{SPC}}}
         }
       }
       child {node {$0.8576$}
         child {node {$0.7406$}
           child {node [rectangle split, rectangle split parts=2, text centered, draw=none] {0.5579\nodepart{second}\textbf{Rate-0}}}
           child {node [rectangle split, rectangle split parts=2, text centered, draw=none] {0.9238\nodepart{second}\textbf{SPC}}}
         }
         child {node [rectangle split, rectangle split parts=2, text centered, draw=none] {0.9753\nodepart{second}\textbf{SPC}}}
       }
     }
     child {node {$0.9508$}
       child {node {$0.9054$}
         child {node [rectangle split, rectangle split parts=2, text centered, draw=none] {0.8230\nodepart{second}\textbf{Type-I}}}
         child {node {$0.9885$}
           child {node [rectangle split, rectangle split parts=2, text centered, draw=none] {0.9774\nodepart{second}\textbf{REP}}}
           child {node [rectangle split, rectangle split parts=2, text centered, draw=none] {0.9998\nodepart{second}\textbf{Rate-1}}}
         }
       }
       child {node {$0.9966$}
         child {node {$0.9933$}
           child {node [rectangle split, rectangle split parts=2, text centered, draw=none] {0.9868\nodepart{second}\textbf{REP}}}
           child {node [rectangle split, rectangle split parts=2, text centered, draw=none] {0.9999\nodepart{second}\textbf{SPC}}}
         }
         child {node [rectangle split, rectangle split parts=2, text centered, draw=none] {1\nodepart{second}\textbf{Rate-1}}}
       }
     };
\end{tikzpicture}
\caption{Tree diagram of the average value of the partially polarized metric functions corresponding to the polarization tree of Fast SCL decoding of PAC$(64,32)$ code using MC rate profile.}
\label{fig: tree-diagram}
\end{figure*}


\section{Metric Polarization Estimation}\label{sec: MetricEstimation}

In this section, we numerically evaluate the path metrics of FSCL decoding for PAC codes across different rate profiles. 
For a PAC$(64, 32)$ code operating at $E_b/N_0 = 2.5~$dB, $\mathbb{E}[\phi(X;Y)] = 0.7944$.
For the Monte Carlo (MC)-based rate profile obtained in \cite{moradi2021monte}, Fig. \ref{fig: tree-diagram} shows the special nodes corresponding to this code. 
A conventional SCL decoder would need to visit $2N - 2 = 126$ nodes of the polarization tree.
However, FSCL significantly reduces this requirement to only $22$ node visits, resulting in more than a 5-fold decrease in the number of node visits.
In this figure, we also show the corresponding average value of the polarized metric functions obtained from the polarization of the mutual information $I(W)$ at each node. 
The leftmost leaf node of the tree is a Rate-0 node and contains 8 LLR values, leading to 8 metric functions, all of which have an average value of $0.1891$. 
On the other hand, the rightmost leaf node shows an average metric function value of $1$, indicating that these are reliable channels, and the decoder is most likely to detect the correct branch for them.

\begin{figure}[htbp] 
\centering
	\includegraphics [width = \columnwidth]{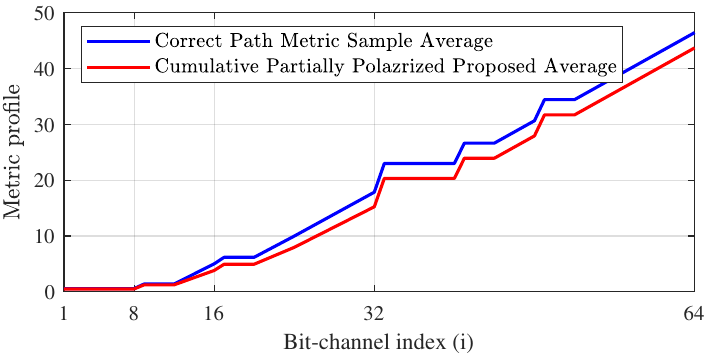}
	\caption{Comparison of the partial optimum average path metric corresponding to the correctly decoded codewords of fast SC decoding of PAC$(64, 32)$ code with the corresponding partially polarized channel capacity rate profile over BI-AWGN at $E_b/N_0 = 2.5~$dB with MC rate profile.} 
	\label{fig: SCMertic_profile_N64_K32_MCRP}
\end{figure}

Fig. \ref{fig: SCMertic_profile_N64_K32_MCRP} shows the sample average path metric values for the FSCL decoding of the PAC$(64, 32)$ code, based on $10^4$ samples of correctly decoded codewords. 
This figure also includes the cumulative values of the average of our proposed polarized metric values derived from the polarization tree shown in Fig. \ref{fig: tree-diagram}.
For the cumulative values, at the $j$th level of decoding, the average of the path metric $\sum_{i=1}^{j}\phi(U_i;\mathbf{Y},\mathbf{U}^{i-1})$ random variable is $\sum_{i=1}^{j}I(W_N^{(i)})$.
It can be observed that the proposed partially polarized average metric provides an excellent estimation of the actual sample average metric function values. 
This suggests that these values can be effectively used for normalizing metric function values in various practical settings as well.

\begin{figure}[htbp] 
\centering
	\includegraphics [width = \columnwidth]{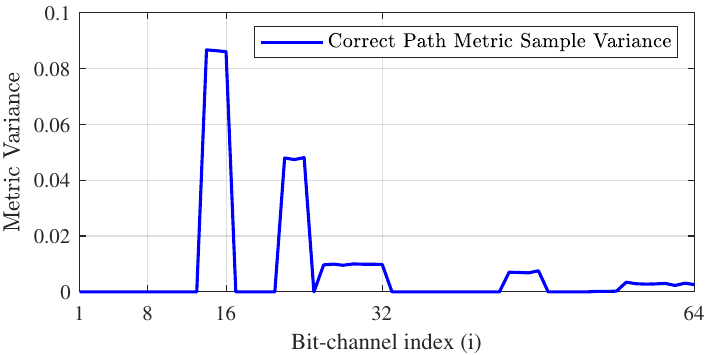}
	\caption{Correct path metric sample variance corresponding to the correctly decoded codewords of fast SC decoding of PAC$(64, 32)$ code of Fig. \ref{fig: SCMertic_profile_N64_K32_MCRP} over BI-AWGN at $E_b/N_0 = 2.5~$dB with MC rate profile.} 
	\label{fig: SCMertic_Var_profile_N64_K32_MCRP}
\end{figure}

Fig. \ref{fig: SCMertic_Var_profile_N64_K32_MCRP} also demonstrates the metric variance of the PAC$(64,32)$ from Fig. \ref{fig: SCMertic_profile_N64_K32_MCRP} for the data bits. 
We observe that the variance at the data bits is nearly zero, suggesting that the average value of the metric function serves as an effective approximation.

\begin{figure}[htbp] 
\centering
	\includegraphics [width = \columnwidth]{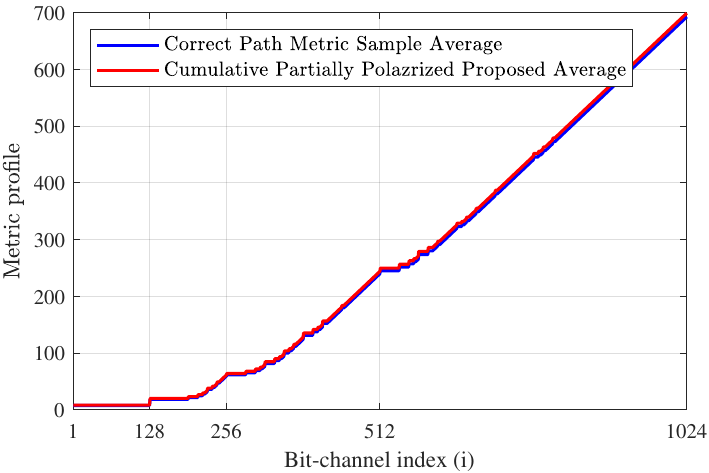}
	\caption{Comparison of the partial optimum average path metric corresponding to the correctly decoded codewords of fast SC decoding of PAC$(1024, 512)$ code with the corresponding partially polarized channel capacity rate profile over BI-AWGN at $E_b/N_0 = 2.5~$dB with GA rate profile.} 
	\label{fig: SCMertic_profile_N1024_K512_PRP}
\end{figure}

\begin{figure}[htbp] 
\centering
	\includegraphics [width = \columnwidth]{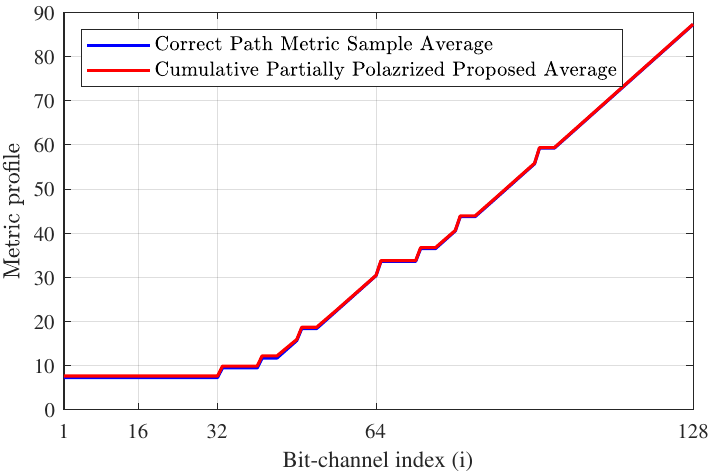}
	\caption{Comparison of the partial optimum average path metric corresponding to the correctly decoded codewords of fast SC decoding of PAC$(128, 64)$ code with the corresponding partially polarized channel capacity rate profile over BI-AWGN at $E_b/N_0 = 2.5~$dB  with GA rate profile.} 
	\label{fig: SCMertic_profile_N128_K64_PRP}
\end{figure}

Figures \ref{fig: SCMertic_profile_N1024_K512_PRP} and \ref{fig: SCMertic_profile_N128_K64_PRP} show the sample averages of the path metric function values for FSCL decoding of the $(1024, 512)$ and $(128, 64)$ PAC codes, respectively. 
These codes use rate profiles obtained by Gaussian approximation.
In these figures, we also plot our proposed average values. 
As the plots illustrate, the average values resulting from our proposed approach provide an excellent match when the rate profile of the code is derived from the polar code construction.

\begin{figure}[htbp]
\centering
	\includegraphics[width=\columnwidth]{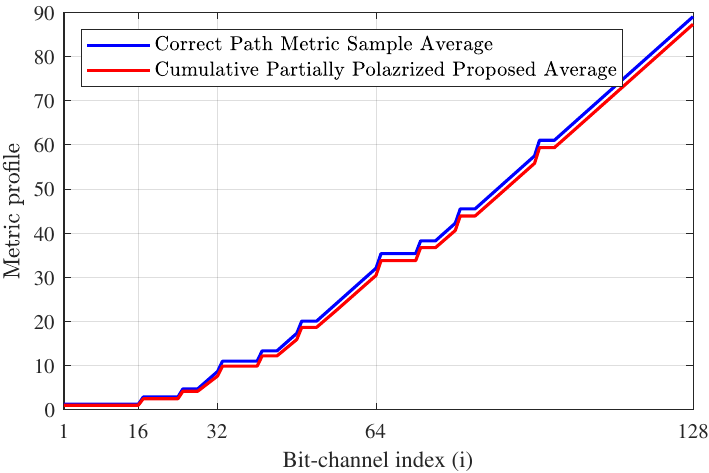}
	\caption{Comparison of the partial average optimum path metric for correctly decoded codewords of fast SC decoding of a PAC$(128, 64)$ code with the corresponding partially polarized channel capacity rate profile over BI-AWGN at $E_b/N_0 = 2.5~$dB, using the RM rate profile.} 
	\label{fig: SCMertic_profile_N128_K64_RMRP}
\end{figure}

Furthermore, Fig. \ref{fig: SCMertic_profile_N128_K64_RMRP} shows the sample average as well as our proposed average of the metric function (obtained from the polarization of $I(W)$) for FSCL decoding of the PAC$(128, 64)$ code, where the rate profile is obtained using RM code construction.
In the next section, we plot the error-correction performance of the different codes studied and the average number of sorting operations required for the FSCL decoding of these codes using our proposed technique.


\section{A List-Pruning Strategy and Numerical Results}\label{sec: NumericalResults}

\subsection{PFSCL}

In this section, we present numerical results for the frame error rate (FER) performance of FSCL decoding for various PAC codes with different code lengths, code rates, and code constructions when a certain list-pruning strategy is deployed based on the proposed metric. 
As demonstrated in the previous section, the path metric values closely follow the average values of our proposed metric in practice.
Utilizing this, whenever splitting is required for the Rate-1 and SPC nodes, we examine the branch metric value. 
If it is lower than a threshold (denoted as $m_T$ in our simulations), we discard that branch. 
For REP or Type-I nodes, we compare the cumulative branch metric values of the chunks at the end of the related segment (for REP nodes, we compare 2 paths; for Type-I nodes, we compare 4 paths). 
If these values are less than the threshold $m_T$, we discard the corresponding path. 
This approach, PFSCL, allows us to avoid sorting if the number of considered paths is less than the list size.

\subsection{VPSCL}
From Theorems \ref{thm:Var} and \ref{thm:UpperBound}, it follows that the value of $m_T$ depends on the polarization of the bit-channels and, consequently, on the rate profile of the code. 
We define
\begin{equation}
    P_{\text{th}} \triangleq \frac{\text{Var}(W_N^{(i)})}{m_i^2},
\end{equation}
and we obtain
\begin{equation}
    m_i = \sqrt{\frac{\text{Var}(W_N^{(i)})}{P_{\text{th}}}}.
\end{equation}
For a given threshold $P_{\text{th}}$, we also propose a varentropy-based list-pruning strategy for the SCL (VPSCL) decoding algorithm to reduce the number of calls to the sorting function.
If the bit-metric function value $\phi(u_i; \mathbf{y}, \mathbf{u}^{i-1})$ is less than $-m_i$, we discard the corresponding path. 
With high probability (depending on $P_{\text{th}}$), this approach retains only the paths whose bit-metric values are close to their average. 
This method is a function of the polarized varentropy and, as a result, works for any rate profile, eliminating the need to tune the parameter $m_T$ as required in previous methods. 
It also suggests that the average number of sorting calls per decoded codeword is approximately equal to the number of bit-channels in $\mathcal{A}$ with high varentropy.

\subsection{Numerical Results}

\begin{figure}[htbp]
\centering
	\includegraphics[width=\columnwidth]{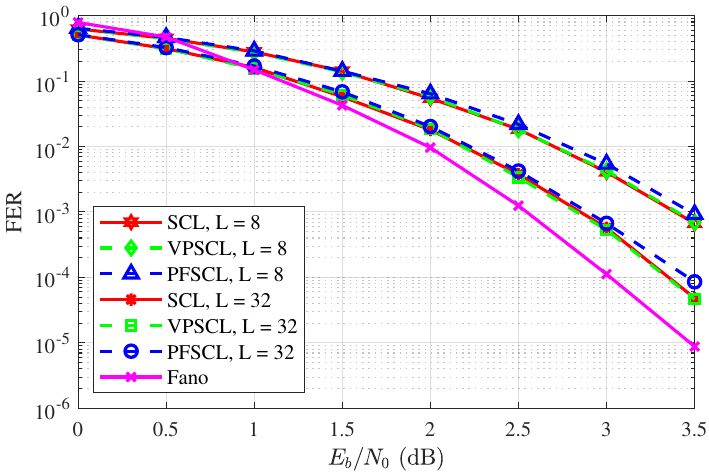}
	\caption{Comparison of the Fano and conventional list decoding with our proposed list pruning of FSCL and VPSCL decoding algorithms for a PAC$(128, 64)$ code when $m_T = -10$ and $P_{th} = 10^{-6}$, respectively}. 
	\label{fig: PFSCLg_pac_listvProposedN128_K64RM}
\end{figure}

Fig. \ref{fig: PFSCLg_pac_listvProposedN128_K64RM} compares the error-correction performance of our proposed PFSCL with conventional SCL decoding for the PAC$(128, 64)$ code using a list size of 8 and 32. 
The PAC code is constructed with the RM rate profile. 
In our proposed method, we prune branches where the corresponding metric function falls below $m_T = -10$. 
PFSCL decoding requires only \(38\) time steps, compared to the \(254\) time steps needed by conventional SCL decoding.
We also include the performance of our proposed VPSCL decoding algorithm with $P_{\text{th}} = 10^{-6}$, corresponding to different list sizes.
Additionally, the performance of the Fano decoding algorithm is presented for comparison in this figure.
In all the numerical results presented in our paper, we use the threshold spacing value $\Delta = 2$ in the Fano decoding algorithm~\cite[Ch. 6]{gallager1968information}, \cite{moradi2024pac}.
Table \ref{table: PAC_N128_K64} shows the sample average number of sorting operations executed by our proposed PFSCL decoder of the PAC$(128, 64)$ code, as plotted in Fig. \ref{fig: PFSCLg_pac_listvProposedN128_K64RM}.
As this table shows, the average number of levels requiring pruning for a high $E_b/N_0$ value of $3.5~$dB is almost half that of a low $E_b/N_0$ value.
The results for our proposed VPSCL decoding algorithm are also included in this table. 
As the results indicate, there is an improvement at lower $E_b/N_0$ values as well. 
We computed the $m_i$ values from the polarized varentropies $V(W_N^{(i)})$ obtained at $E_b/N_0 = 2.5~$dB.

\begin{table*}[htbp]
\footnotesize
\centering
\caption{Sample average number of sorting operations for our proposed PFSCL and VPSCL decoder algorithms for the PAC$(128, 64)$ code as plotted in Fig. \ref{fig: PFSCLg_pac_listvProposedN128_K64RM}.}
\renewcommand{\arraystretch}{1.2}
\begin{tabular}{ccccccccc}
\hline
$E_b/N_0$ {[}dB{]}          & 0.0    & 0.5    & 1.0    & 1.5    & 2.0    & 2.5    & 3.0   & 3.5    \\ \hline
\# of sorting ($L = 32$, $m_{T} = -10$) & 65.93 & 64.53 & 61.79 & 56.60 & 49.89 & 43.04 & 37.29 & 32.78 \\ \hline
\# of sorting ($L = 8$, $m_{T} = -10$) & 61.53 & 58.69 & 54.20 & 48.14 & 41.44 & 34.61 & 28.38 & 23.43 \\ \hline
\# of sorting ($L = 32$, $P_{th} = 10^{-6}$) & 32.10 & 32.01 & 32.06 & 32.09 & 32.15 & 32.21 & 32.22 & 32.32 \\ \hline
\# of sorting ($L = 8$, $P_{th} = 10^{-6}$) & 32.68 & 32.46 & 32.25 & 31.94 & 31.62 & 31.27 & 31.00 & 30.83 \\ \hline
\end{tabular}\label{table: PAC_N128_K64}
\end{table*}

\begin{figure}[htbp]
\centering
	\includegraphics[width=\columnwidth]{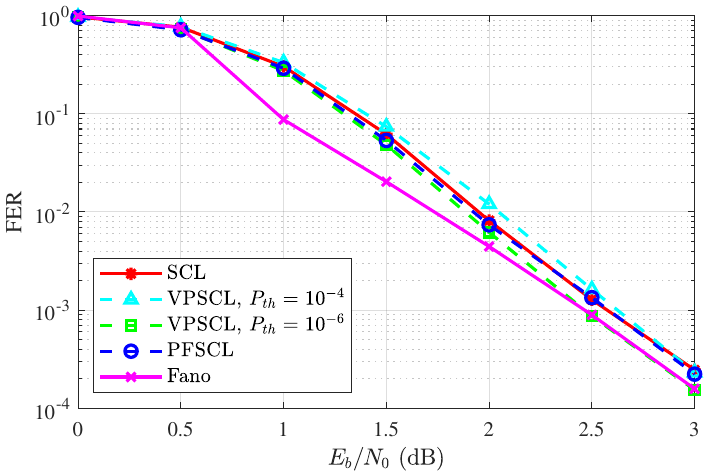}
	\caption{Comparison of the Fano and conventional list decoding with our proposed pruning of FSCL with $m_T = -10$ and VPSCL decoding algorithms with {$P_{th}=10^{-4}$ and $10^{-6}$} for a PAC$(1024, 512)$ code all with a list size of $4$.}
	\label{fig: PFSCLg_pac_listvProposedN1024_K512RM}
\end{figure}

Fig. \ref{fig: PFSCLg_pac_listvProposedN1024_K512RM} compares the error-correction performance of our proposed PFSCL with conventional SCL decoding for the PAC$(1024, 512)$ code using a list size of $4$. 
The code is constructed with the Gaussian approximation rate profile at $E_b/N_0 = 2.5~$dB value. 
In our proposed method, we prune branches where the corresponding metric function falls below $m_T = -10$.
PFSCL decoding requires only \(162\) time steps, compared to the \(2046\) time steps needed by conventional SCL decoding.
We also include the performance of our proposed VPSCL decoding algorithm with $P_{\text{th}} = 10^{-6}$ and $P_{\text{th}} = 10^{-4}$.
Additionally, the performance of the Fano decoding algorithm is presented for comparison in this figure.
Table \ref{table: PAC_N1024_K512} shows the sample average number of sorting operations executed by our proposed PFSCL decoder of the PAC$(1024, 512)$ code, as plotted in Fig. \ref{fig: PFSCLg_pac_listvProposedN1024_K512RM}.
As this table shows, the average number of levels requiring sorting for a high $E_b/N_0$ value of $3~$dB is about $5\%$ of a conventional SCL decoding algorithm.
The results for our proposed VPSCL decoding algorithm are also included in this table. 
As the results indicate, there is an improvement at lower $E_b/N_0$ values as well. 
We computed the $m_i$ values from the polarized varentropies $V(W_N^{(i)})$ obtained at $E_b/N_0 = 2.5~$dB.

\begin{table*}[htbp]
\footnotesize
\centering
\caption{Sample average number of sorting operations for our proposed FSCL and VPSCL decoder algorithms for the PAC$(1024, 512)$ code as plotted in Fig. \ref{fig: PFSCLg_pac_listvProposedN1024_K512RM}.}
\renewcommand{\arraystretch}{1.2}
\begin{tabular}{cccccccc}
\hline
$E_b/N_0$ {[}dB{]}          & 0.0    & 0.5    & 1.0    & 1.5    & 2.0    & 2.5    & 3.0      \\ \hline
\# of sorting ($L = 4$, $m_{T} = -10$) & 298.69 & 262.75 & 204.79 & 131.53 & 67.02 & 35.78 & 28.04 \\ \hline
\# of sorting ($L = 4$, $P_{th} = 10^{-6}$) & 114.33 & 111.43 & 108.00 & 104.71 & 100.81 & 96.20 & 89.79 \\ \hline
\# of sorting ($L = 4$, $P_{th} = 10^{-4}$) & 73.45 & 66.02 & 55.50 & 41.83 & 37.10 & 40.85 & 41.19 \\ \hline
\end{tabular}\label{table: PAC_N1024_K512}
\end{table*}

\begin{figure}[htbp]
\centering
	\includegraphics[width=\columnwidth]{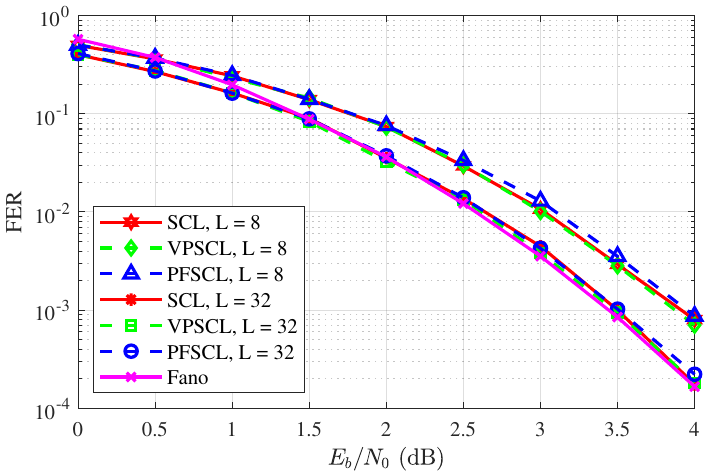}
	\caption{Comparison of the Fano and conventional list decoding with our proposed pruning of FSCL and VPSCL decoding algorithms for a PAC$(64, 32)$ code when $m_T = -10$ and $P_{th} = 10^{-6}$, respectively}. 
	\label{fig: PFSCLg_pac_listvProposedN64_K32MC}
\end{figure}

Fig. \ref{fig: PFSCLg_pac_listvProposedN64_K32MC} compares the error-correction performance of our proposed PFSCL with conventional SCL decoding for the PAC$(64, 32)$ code using a list size of 8 and $32$. 
The PAC code is constructed with the MC rate profile at $E_b/N_0 = 5~$dB value \cite{moradi2021monte}. 
In our proposed method, we prune branches where the corresponding metric function falls below $m_T = -10$. 
PFSCL decoding requires only \(22\) time steps, compared to the \(126\) time steps needed by conventional SCL decoding.
We also include the performance of our proposed VPSCL decoding algorithm with $P_{\text{th}} = 10^{-6}$, corresponding to different list sizes.
Additionally, the performance of the Fano decoding algorithm is presented for comparison in this figure.
Table \ref{table: PAC_N64_K32} shows the sample average number of sorting operations executed by our proposed PFSCL decoder of the PAC$(64, 32)$ code, as plotted in Fig. \ref{fig: PFSCLg_pac_listvProposedN64_K32MC}.
The results for our proposed VPSCL decoding algorithm are also included in this table. 
We computed the $m_i$ values from the polarized varentropies $V(W_N^{(i)})$ obtained at $E_b/N_0 = 2.5~$dB.

\begin{table*}[htbp]
\footnotesize
\centering
\caption{Sample average number of sorting operations for our proposed PFSCL and VPSCL decoder algorithms for the PAC$(64, 32)$ code as plotted in Fig. \ref{fig: PFSCLg_pac_listvProposedN64_K32MC}.}
\renewcommand{\arraystretch}{1.2}
\begin{tabular}{cccccccccc}
\hline
$E_b/N_0$ {[}dB{]}          & 0.0    & 0.5    & 1.0    & 1.5    & 2.0    & 2.5    & 3.0    & 3.5    & 4.0      \\ \hline
\# of sorting ($L = 32$, $m_{T} = -10$) & 29.27 & 29.11 & 28.89 & 28.60 & 28.18 & 27.56 & 26.55 & 24.93 & 22.55 \\ \hline
\# of sorting ($L = 8$, $m_{T} = -10$) & 29.16 & 28.67 & 28.02 & 27.18 & 25.85 & 23.81 & 21.06 & 18.00 & 14.87 \\ \hline
\# of sorting ($L = 32$, $P_{th} = 10^{-6}$) & 18.91 & 18.87 & 18.85 & 18.82 & 18.79 & 18.75 & 18.68 & 18.61 & 18.49 \\ \hline
\# of sorting ($L = 8$, $P_{th} = 10^{-6}$) & 19.86 & 19.81 & 19.76 & 19.72 & 19.67 & 19.59 & 19.52 & 19.42 & 19.24 \\ \hline
\end{tabular}\label{table: PAC_N64_K32}
\end{table*}

\begin{figure}[htbp]
\centering
	\includegraphics[width=\columnwidth]{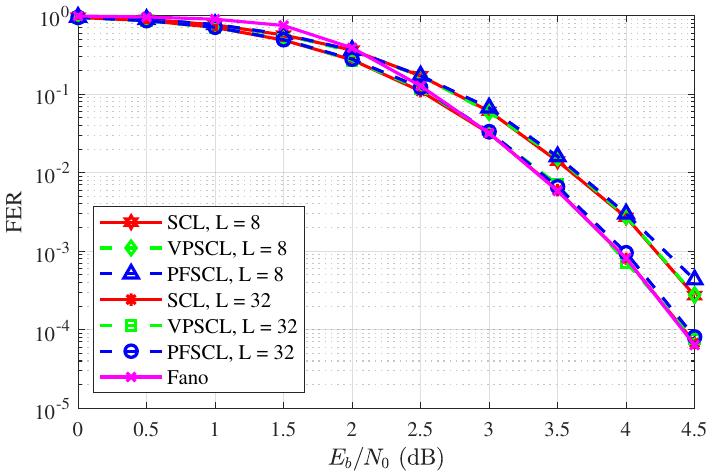}
	\caption{Comparison of the Fano and conventional list decoding with our proposed pruning of FSCL decoding algorithm for a PAC$(128, 99)$ code when $m_T = -15$.} 
	\label{fig: PFSCLg_pac_listvProposedN128_K99RM}
\end{figure}

Fig. \ref{fig: PFSCLg_pac_listvProposedN128_K99RM} compares the error-correction performance of our proposed PFSCL with conventional SCL decoding for the PAC$(128, 99)$ code using a list size of $8$ and $32$. 
The PAC code is constructed with the RM rate profile. 
In our proposed method, we prune branches where the corresponding metric function falls below $m_T = -15$.
PFSCL decoding requires only \(28\) time steps, compared to the \(254\) time steps needed by conventional SCL decoding.
We also include the performance of our proposed VPSCL decoding algorithm with $P_{\text{th}} = 10^{-6}$, corresponding to different list sizes.
Additionally, the performance of the Fano decoding algorithm is presented for comparison in this figure.
Table \ref{table: PAC_N128_K99} shows the sample average number of sorting operations executed by our proposed PFSCL decoder of the PAC$(128, 99)$ code, as plotted in Fig. \ref{fig: PFSCLg_pac_listvProposedN128_K99RM}.
As this table shows, the average number of levels requiring pruning for a high $E_b/N_0$ value of $5~$dB is about $33\%$ of a conventional SCL decoding algorithm and the number of node visits on the polarization tree is $28$ while for the conventional SCL decoding it is $254$.
The results for our proposed VPSCL decoding algorithm are also included in this table. 
As the results indicate, there is an improvement at lower $E_b/N_0$ values as well. 
We computed the $m_i$ values from the polarized varentropies $V(W_N^{(i)})$ obtained at $E_b/N_0 = 3.5~$dB.

\begin{table*}[htbp]
\footnotesize
\centering
\caption{Sample average number of sorting operations for our proposed PFSCL and VPSCL decoder algorithms for the PAC$(128, 99)$ code as plotted in Fig. \ref{fig: PFSCLg_pac_listvProposedN128_K99RM}.}
\renewcommand{\arraystretch}{1.2}
\begin{tabular}{ccccccccccc}
\hline
$E_b/N_0$ {[}dB{]}          & 0.0    & 0.5    & 1.0    & 1.5    & 2.0    & 2.5    & 3.0    & 3.5    & 4.0 & 4.5      \\ \hline
\# of sorting ($L = 32$, $m_{T} = -15$) & 98.31 & 96.50 & 94.77 & 92.93 & 90.32 & 85.28 & 76.44 & 64.45 & 52.76 & 42.66 \\ \hline
\# of sorting ($L = 8$, $m_{T} = -15$) & 98.28 & 94.82 & 90.69 & 85.75 & 79.36 & 70.96 & 60.51 & 49.24 & 38.94 & 30.25 \\ \hline
\# of sorting ($L = 32$, $P_{th} = 10^{-6}$) & 36.11 & 35.56 & 35.13 & 34.76 & 34.57 & 34.44 & 34.51 & 34.58 & 34.66 & 34.65 \\ \hline
\# of sorting ($L = 8$, $P_{th} = 10^{-6}$) & 36.56 & 35.97 & 35.39 & 34.86 & 34.34 & 33.90 & 33.51 & 33.21 & 32.96 & 32.66 \\ \hline
\end{tabular}\label{table: PAC_N128_K99}
\end{table*}

\begin{figure}[htbp]
\centering
	\includegraphics[width=\columnwidth]{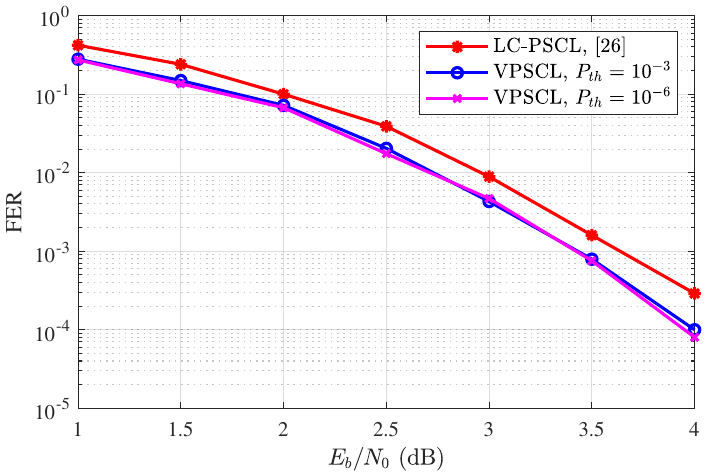}
	\caption{Comparison of our proposed VPSCL decoding algorithm with different \( P_{th} \) values for a PAC$(128,64)$ code against the LC-PSCL decoding algorithm introduced in \cite{yao2024low} for a CRC-aided Polar$(128,64)$ code, both with a list size of $8$.
} 
	\label{fig: VPSCLg_pac_v_Yao_N128_K64_L8}
\end{figure}

\begin{figure}[htbp]
\centering
	\includegraphics[width=\columnwidth]{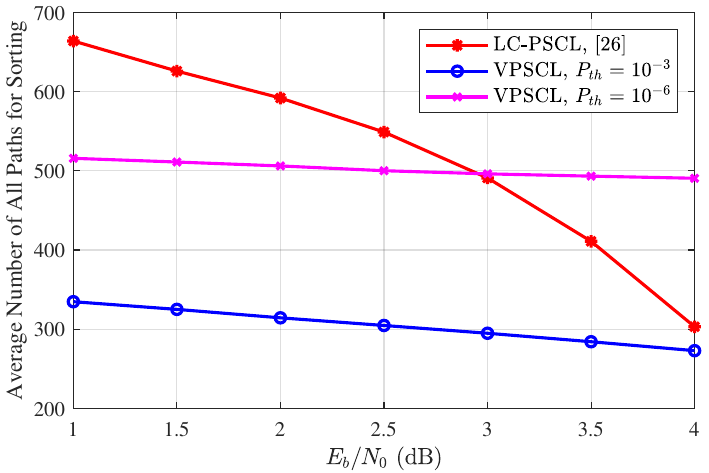}
	\caption{Comparison of the sample average number of all paths requiring sorting in our proposed VPSCL decoding algorithm with different \( P_{th} \) values for a PAC(128, 64) code, against the LC-PSCL decoding algorithm introduced in \cite{yao2024low} for a CRC-aided Polar(128, 64) code, both with a list size of 8.
} 
	\label{fig: AllSORTCount_pac_v_Yao_N128_K64_L8}
\end{figure}

\begin{figure}[htbp]
\centering
	\includegraphics[width=\columnwidth]{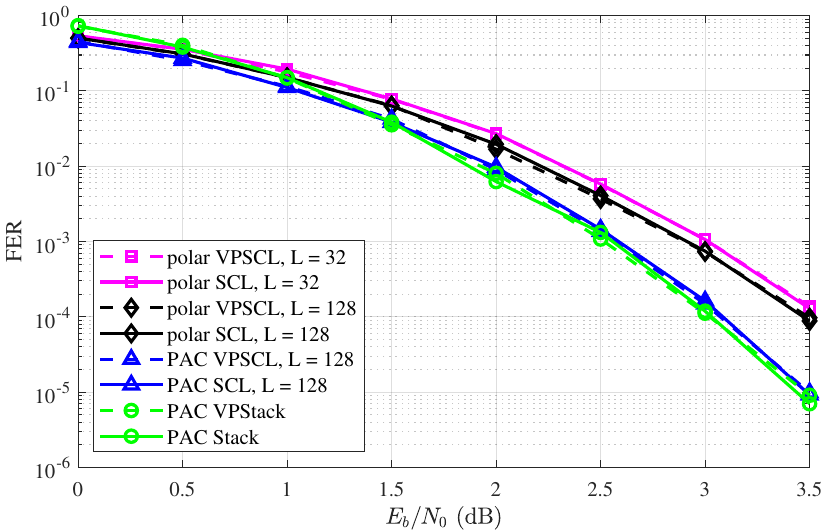}
	\caption{Comparison of conventional stack and list decoding with our proposed pruning strategies—VPSCL and VPStack—for polar and PAC $(128, 64)$ codes. For VPSCL, we use $P_{\text{th}} = 10^{-6}$, and for VPStack, we use $P_{\text{th}} = 10^{-4}$.
} 
	\label{fig: PFSCLg_pac_polar_listvStack_ProposedN128_K64RM}
\end{figure}

\begin{table*}[htbp]
\footnotesize
\centering
\caption{Sample average number of sorting operations for our proposed VPSCL decoder algorithm for the PAC and polar$(128, 64)$ codes as plotted in Fig. \ref{fig: PFSCLg_pac_polar_listvStack_ProposedN128_K64RM}.}
\renewcommand{\arraystretch}{1.2}
\begin{tabular}{ccccccccc}
\hline
$E_b/N_0$ {[}dB{]}          & 0.0    & 0.5    & 1.0    & 1.5    & 2.0    & 2.5    & 3.0    & 3.5        \\ \hline
\# of sorting (polar, $L = 32$, $P_{th} = 10^{-6}$) & 30.87 & 30.54 & 30.13 & 29.61 & 29.12 & 28.57 & 28.09 & 27.68 \\ \hline
\# of sorting (polar, $L = 128$, $P_{th} = 10^{-6}$) & 30.14 & 30.00 & 29.84 & 29.68 & 29.48 & 29.27 & 29.06 & 28.85 \\ \hline
\# of sorting (PAC, $L = 128$, $P_{th} = 10^{-6}$) & 30.46 & 30.48 & 30.56 & 30.70 & 30.91 & 31.19 & 31.47 & 31.74 \\ \hline
\end{tabular}\label{table: PACvsPolar_N128_K64}
\end{table*}

\begin{figure}[htbp]
\centering
	\includegraphics[width=\columnwidth]{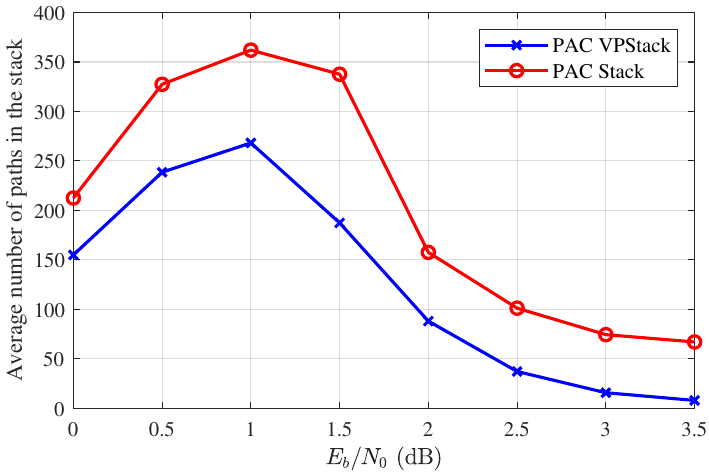}
	\caption{Comparison of the average number of paths explored by conventional stack decoding and our proposed VPSCL pruning strategy for the PAC $(128, 64)$ code with $P_{\text{th}} = 10^{-4}$.
} 
	\label{fig: VPStackUsed_pac_N128_K64RM}
\end{figure}

Fig. \ref{fig: VPSCLg_pac_v_Yao_N128_K64_L8} compares the error-correction performance of our proposed VPSCL decoding algorithm for a PAC(128, 64) code with the pruning algorithm introduced in \cite{yao2024low} for a CRC-aided Polar(128, 64) code. 
For VPSCL algorithm the \( m_i \) values are computed from the polarized varentropies \( V(W_N^{(i)}) \) obtained at \( E_b/N_0 = 2.5~\text{dB} \). 
In this example, the PAC code demonstrates better performance compared to the polar code.

Fig. \ref{fig: AllSORTCount_pac_v_Yao_N128_K64_L8} shows the sample average number of all paths requiring sorting operations by our proposed VPSCL decoder for the PAC(128, 64) code versus the algorithm introduced in \cite{yao2024low}, corresponding to Fig. \ref{fig: VPSCLg_pac_v_Yao_N128_K64_L8}. As mentioned earlier, our proposed polarized variance-based algorithm depends on the number of bit channels that are not fully polarized, leading to a nearly constant number of paths requiring sorting across different \( E_b/N_0 \) values. This behavior provides an advantage, especially at lower \( E_b/N_0 \) values.

In Fig.~\ref{fig: PFSCLg_pac_polar_listvStack_ProposedN128_K64RM}, we compare the performance of VPSCL and SCL decoding for polar and PAC codes with various list sizes, alongside the performance of VPStack and conventional stack decoding for PAC codes.  
As shown in the figure, PAC codes decoded with list decoding using a list size of 128 can achieve performance comparable to that of sequential decoding (stack decoding) of PAC codes.
Table \ref{table: PACvsPolar_N128_K64} also shows the sample average number sorting operations executed by our proposed VPSCL decoder of the polar and PAC $(128,64)$ codes correpsinding to Fig. \ref{fig: PFSCLg_pac_polar_listvStack_ProposedN128_K64RM}.
Although PAC codes has a much better coding gain compared to the polar codes, the average number of levels requiring sorting for both polar and PAC codes are the same indicating that this depends on the number of bit-channels with a high varentropy.
Fig.~\ref{fig: VPStackUsed_pac_N128_K64RM} also compares the average number of paths in the stack for our proposed VPStack algorithm and the conventional stack decoding algorithm, corresponding to the performance results shown in Fig.~\ref{fig: PFSCLg_pac_polar_listvStack_ProposedN128_K64RM}.

\section{Conclusion}\label{sec: Conclusion}
In this paper, we introduced the concept of metric function polarization and proved that the average and variance of the polarized metric function variables at each step of polarization are equal to the corresponding polarized mutual information and polarized channel varentropy. We also demonstrated that, on average, the value of the partially polarized metric function for an incorrect branch is negative.
Furthermore, we established that the probability of the bit-metric value for a correct path diverging from the bit-metric mutual information is upper bounded by a fraction where the numerator contains the bit-channel varentropy, and the denominator represents the magnitude of divergence. Based on this upper bound, we proposed a list-pruning strategy for the SCL decoding algorithm that leverages polarized varentropy. This technique avoids unnecessary calls to the sorting function when the bit-channel is well polarized.
Additionally, we introduced a constant threshold for pruning paths whose metric values are significantly diverged from the bit-channel mutual information and applied this approach to the fast SCL decoding algorithm. As demonstrated by our numerical results, the proposed algorithm significantly reduces the number of sorting operations, especially at high \(E_b/N_0\) values.


\ifCLASSOPTIONcaptionsoff
  \newpage
\fi

\bibliographystyle{IEEEtran}
\bibliography{bibliography}

\end{document}